\documentclass[12pt]{article}

\usepackage[T1]{fontenc}
\usepackage[utf8]{inputenc}

\usepackage[margin=1in]{geometry}
\usepackage[protrusion=true,expansion=false]{microtype}
\usepackage{amsmath,amssymb,amsthm}
\usepackage{bbm}
\usepackage{graphicx}
\usepackage{booktabs}
\usepackage{natbib}
\usepackage{enumitem}
\usepackage{tikz}
\usepackage{setspace}
\usepackage{xurl}
\usepackage[colorlinks=true,linkcolor=blue,citecolor=blue,urlcolor=blue]{hyperref}

\usetikzlibrary{arrows.meta,positioning}

\newtheorem{theorem}{Theorem}
\newtheorem{proposition}{Proposition}
\newtheorem{lemma}{Lemma}
\newtheorem{corollary}{Corollary}
\newtheorem{assumption}{Assumption}
\newtheorem{definition}{Definition}
\newtheorem{remark}{Remark}
\newtheorem{example}{Example}

\newcommand{\E}{\mathbb{E}}
\newcommand{\Prob}{\mathbb{P}}
\newcommand{\R}{\mathbb{R}}

\newcommand{\Var}{\operatorname{Var}}
\newcommand{\Cov}{\operatorname{Cov}}

\title{Reputation and Disclosure in Dynamic Networks\thanks{I thank Bruno Strulovici for helpful comments. Ben Golub and Yann Calvó López's \href{https://refine.ink}{Refine.ink} was used to audit notation for consistency and proofs for clarity. Any remaining errors are my own. A \href{https://www.sebastianbuhai.com/papers/publications/reputation_disclosure_networks_OA.pdf}{supplementary online appendix} contains additional worked benchmarks, broader extensions, and institutional notes.}}
\author{I.\ Sebastian Buhai\thanks{Contact email: \texttt{sebastian.buhai@sofi.su.se}. Website: \href{https://www.sebastianbuhai.com}{sebastianbuhai.com}}\\
{SOFI at Stockholm University}\\
{Instituto de Economia at UC Chile}\\
{NIPE at Minho University}}
\date{First version: December 28, 2025. This version: May 24, 2026. \href{https://www.sebastianbuhai.com/papers/publications/reputation_disclosure_networks.pdf}{Latest version}.}

\begin{document}
\hypersetup{pdftitle={Reputation and Disclosure in Dynamic Networks},pdfauthor={I. Sebastian Buhai},pdfsubject={Docketed hard evidence, informative retention, and common-record censoring},pdfkeywords={hard evidence, verifiable disclosure, informative retention, public review, dynamic disclosure, reputation, network formation, common-record censoring}}
\maketitle

\begin{abstract}

Public delay can be informative when the existence, custodian, and review dates of hard evidence are observed. I study a disclosure protocol in which a sealed record is docketed, held by a public custodian, and revealed only at terminal disclosure. At each review, retention is not silence: it rules out the states in which the holder would have relayed or disclosed. This censoring event yields an exact Bayesian filter. Under interval strategies, the public posterior is summarized by finitely many support endpoints. A compact reputation benchmark verifies Markov perfect Bayesian equilibria with such strategies and gives finite time resolution on compact unresolved slices. With two certified routes for the same record, retention on one route changes what remains feasible on the other before it acts. Common record censoring creates network value that pairwise formation can miss.

\bigskip

\end{abstract}

\section{Introduction}\label{sec:intro}

\subsection{Question, object, and contribution}

Many economic records are not disclosed as soon as they are created. They sit in audit queues, compliance files, review systems, and certification chains where outsiders know that a file exists, who holds it, and when review occurs. In such settings, delay is not missing information. If a custodian could have advanced the file at a public review date and did not, the choice reveals which states remain possible.

I study this logic in a dynamic disclosure model with hard evidence. A sealed record is dated and publicly docketed. Custody and review dates are public. Content becomes public only at terminal disclosure. At a review date, a holder either retains, relays, or discloses. Retention is a censoring event: it rules out the states in which the holder would have relayed or disclosed. The public signal supplies a Gaussian reference likelihood, while the review history restricts its support. Under interval policies, the exact posterior is summarized by finitely many endpoints.

The model separates custody from audit information. Custody determines which sealed file states remain possible. The audit record moves reputation and governs future influence, assignments, or access. This separation gives a single censored evidence posterior while review behavior disciplines the intermediary. The main benchmark keeps custody regions common across audit types, giving a one dimensional single kernel filter. When custody choices depend on audit type, the filter leads to weighted supports, as shown in the supplementary online appendix.

My first contribution is the exact censoring recursion. Public retention can be informative even when the record remains sealed. Without public custody and review dates, delay is a mixture over latent review histories rather than a finite endpoint state. The second contribution is a compact equilibrium benchmark. In an Ornstein Uhlenbeck quadratic environment, primitive affine classes verify interval policy Markov perfect Bayesian equilibria. The same discipline yields finite time resolution of old records on compact unresolved slices, with Gaussian tail bounds beyond compact support. Finally, the paper's third contribution is the network implication. When two certified routes carry the same record, retention on one route changes what is feasible on the other before it acts. Common record censoring adds planner value that private bilateral formation need not price.

A canonical institution is an internal audit or compliance docket. A case is opened and logged, its contents are protected, the case owner is public, review dates are scheduled, retention is recorded, escalation corresponds to relay, final release corresponds to terminal disclosure, and review behavior affects future assignments. In the paper, I formalize this protocol; I discuss related institutions in the supplementary online appendix.

\subsection{Relation to the literature and plan}

The paper adds public custody and public review of an already docketed sealed record to hard information disclosure. It builds on unraveling and verifiable evidence models
\citep{Grossman1981Disclosure,Milgrom1981GoodNews,Verrecchia1983,Dye1985,JungKwon1988,SeidmannWinter1997,BullWatson2004,Dziuda2011,HagenbachKoesslerPerezRichet2014,HartKremerPerry2017,BenPorathDekelLipman2019,Rappoport2025},
dynamic disclosure and information flow models
\citep{AcharyaDeMarzoKremer2011,GuttmanKremerSkrzypacz2014,OrlovSkrzypaczZryumov2020,LeeMarinovic2025},
and intermediary, certification, evidence acquisition, and reputation models
\citep{Lizzeri1999,ElyValimaki2003,Ivanov2010,BoardMeyerTerVehn2013,CheSeverinov2017,MarinovicSkrzypaczVaras2018,Herresthal2022,Arrora2026,Shishkin2026}.
Relative to disclosure networks
\citep{JacksonWolinsky1996,GaleottiGhiglinoSquintani2013,SquintaniStrategicDisclosureNetworks,BaumannDutta2026},
the externality is dynamic common record censoring.

Section~\ref{sec:model} gives the protocol and equilibrium concept. Section~\ref{sec:static} records the static benchmark. Section~\ref{sec:dynamic} proves the exact filter, compact benchmark, and old record resolution results. Section~\ref{sec:network} studies common record censoring and formation. Section~\ref{sec:conclusion} concludes. The manuscript appendix contains the core proofs and filtering formulas; the online appendix contains worked benchmarks, extensions, and institutional notes.

\section{Environment}\label{sec:model}

\subsection{Players, fundamentals, and public information}\label{subsec:model_players}

Time is continuous, $t\in[0,\infty)$, and uncertainty is defined on a filtered probability space $(\Omega,\mathcal F,(\mathcal F_t)_{t\ge 0},\Prob)$ satisfying the usual conditions. The payoff-relevant fundamental $(X_t)_{t\ge 0}$ follows the It\^o diffusion
\begin{equation}
  dX_t = \mu(X_t)\,dt + \sigma(X_t)\,dB_t,
  \label{eq:fundamental_diffusion}
\end{equation}
where $(B_t)_{t\ge 0}$ is a standard Brownian motion and $\mu:\R\to\R$ and $\sigma:\R\to(0,\infty)$ are globally Lipschitz. Later sections specialize to an Ornstein-Uhlenbeck benchmark. All agents also observe the public signal
\begin{equation}
  dY_t = hX_t\,dt + \sigma_Y\,dW_t^Y,
  \label{eq:public_signal}
\end{equation}
with $h\ge0$, $\sigma_Y>0$, and $(W_t^Y)_{t\ge 0}$ a standard Brownian motion independent of $B$; $h=0$ is the no-learning benchmark. The exact public posterior conditions jointly on the signal path and public censoring events; the uncensored filter below is the likelihood coordinate that makes this factorization explicit.

The benchmark protocol has one publicly docketed sealed record, public custody logs, posted review calendars, publicly observed review dates, a distinct public audit mark attached to review outcomes, and no supersession while the record remains unresolved. The canonical interpretation is a compliance or internal-audit docket:
\begin{center}\small
\begin{tabular}{ll}
\toprule
Model object & Institutional counterpart\\
\midrule
Known dated record & opened case or audit file\\
Sealed content & protected evidence\\
Public custody & logged case owner\\
Public review date & scheduled oversight review\\
Retention & file remains with custodian\\
Relay / terminal disclosure & escalation / final release\\
Audit reputation & future assignment, access, or influence\\
Certified copies & parallel review channels holding the same file\\
\bottomrule
\end{tabular}
\end{center}
The docket makes silence public, custody identifies the hidden holder gap, review dates make retention a censoring event, age is observed, and the audit record updates reputation.

The set of agents is $N=\{0,1,\dots,n\}$. Agent $0$ is the decision maker, who chooses $y_t\in\R$. The remaining agents are senders; a subset $E\subseteq N\setminus\{0\}$ are experts and the rest are relay intermediaries. This distinction is topological: experts are the source nodes that originate docketed records, while all senders have the same payoff form once they hold a record. The decision maker's flow payoff is
\[
  u_0(y_t,X_t)=-(y_t-X_t)^2.
\]
Each sender $i\in N\setminus\{0\}$ has publicly known directional bias $b_i\in\R$ and flow payoff
\begin{equation}
  u_i(y_t,X_t,R_t^i)
  =
  -\alpha_i\bigl(y_t-X_t-b_i\bigr)^2 + \beta_i g_i(R_t^i),
  \label{eq:flow_payoff_i}
\end{equation}
where $\alpha_i>0$, $\beta_i>0$, $R_t^i\in\mathcal R_i\subseteq\R$ is a public reputational state, and $g_i$ is continuous and strictly increasing. Review outcomes move $R_t^i$ through public transition maps generated by an audit mark separate from sealed-evidence inference. The maps may be primitive or generated by the audit experiment in Appendix~C; the public updates audit reputation on the mark and evidence beliefs on the custody event.

All players maximize expected discounted payoffs with a common discount rate $\rho>0$. Starting from any public date $t_0$, agent $i$ evaluates a continuation strategy by
\[
  \E\!\left[\int_{t_0}^{\infty} e^{-\rho(t-t_0)}
  u_i(y_t,X_t,R_t^i)\,dt\,\middle|\,\mathcal F_{t_0}^i\right],
\]
with the convention that the decision maker uses $u_0(y_t,X_t)$ and no reputational coordinate.

\subsection{Network and paths of communication}\label{subsec:model_network}

A directed communication network is a graph $G=(N,L)$ with link set $L\subseteq N\times N$. If $(i,j)\in L$, agent $i$ can send verifiable records to agent $j$. Now write $\mathcal N_i^+=\{j:(i,j)\in L\}$ and $\mathcal N_i^- = \{j:(j,i)\in L\}$ for out- and in-neighbors. A directed expert-to-decision-maker (expert-to-DM) path is any finite sequence of distinct nodes
\[
  P=(v_0,v_1,\dots,v_m)
\]
with $v_0=e\in E$, $v_m=0$, and $(v_r,v_{r+1})\in L$ for each $r<m$; also write $P=(e,i_1,\dots,i_k,0)$.

Sections~\ref{sec:dynamic} and~\ref{sec:network} below will use two benchmark architectures: a fixed path $P$, and the two-route race that compares the serial line $e\to i\to j\to 0$ with the overlapping routes $e\to i\to 0$ and $e\to j\to 0$.

\subsection{Information structure and disclosure technology}\label{subsec:model_info}

As part of the main analysis, I study one certified record
\[
  \varrho_\tau=(X_\tau,\tau)
\]
whose existence is publicly docketed at date $\tau$ while its content remains sealed. The original timestamp, current custody, relay receipts, posted review calendars, realized review dates, and any terminal disclosure are public. If $C_t=i\neq\varnothing$ is the current custodian, sender $i$ privately observes the sealed content and forms the posterior mean
\[
  \mu_t^i := \E[X_t\mid X_\tau,\mathcal F_t^p],
\qquad
  d_t^i := \mu_t^i-m_t,
\]
where $m_t=\E[X_t\mid \mathcal F_t^p]$ is the public posterior mean. The hidden object is the scalar gap between the holder's posterior and the public belief.

If intermediary $i$ relays the operative record to a non-decision-maker neighbor $j\neq 0$, the public observes a certified custody transfer together with the original timestamp, but the content remains hidden. If the record is sent to the decision maker, the content, timestamp, and terminal discloser become public. In the two-route race, the expert may place certified copies of the same docketed record on both branches. Because those copies inherit the same content and timestamp, the hidden state remains one common dated record gap until the first terminal disclosure.

For each sender $i\in N\setminus\{0\}$, let $(\lambda_t^i)_{t\ge 0}$ be an $\mathcal F^p$-adapted c\`adl\`ag review calendar process taking values in $\{0,\bar\lambda_i\}$. The value $\bar\lambda_i$ means that the channel is active, while the value $0$ records a posted pause. The benchmark takes this calendar to be fixed ex ante or generated by the public state, and in either case independent of the sealed content; its left limit $\lambda_{t-}^i$ is the predictable review intensity. When arrival rates are allowed to depend on the sealed content, the ocurrence of a review itself carries a likelihood term before the retention update; I treat that variant in the supplementary online appendix (Online Appendix OB).

Given the calendar path, realized review dates are generated by a public point process $(N_t^i)_{t\ge 0}$ with compensator
\[
  \Lambda_t^i = \int_0^t \lambda_{s-}^i\,ds.
\]
A jump of $N^i$ is a public review mark. If $i$ currently holds the record, outsiders then observe relay, terminal disclosure, or explicit retention. Conditional on the public calendar path and custody history, no-jump events between review dates induce no censoring update.

The public filtration $(\mathcal F_t^p)_{t\ge 0}$ is generated by the public signal path, docketing events, relay receipts, realized review dates, terminal disclosures, calendar paths, and the decision maker's past actions. From the state dictionary onward, the main results use the Ornstein-Uhlenbeck linear Gaussian specialization anticipated above. Under that specialization, the exact state dictionary is then
\[
  m_t = \E[X_t\mid \mathcal F_t^p],
  \qquad
  \tilde v_t = v_t^{\mathrm{ref}},
\]
where $v_t^{\mathrm{ref}}$ is the variance from the uncensored Ornstein-Uhlenbeck public signal reference filter, and
\[
\begin{aligned}
  d_t^i&=\mu_t^i-m_t,\\
  (\bar d_t,q_t)&=\text{Gaussian reference kernel parameters for } d_t^{C_t},\\
  K_t&=\text{its feasible support}.
\end{aligned}
\]
The distinction between $\tilde v_t$ and $(\bar d_t,q_t,K_t)$ is central. Once review date censoring occurs, the post-censoring moments are recovered by conditioning the Gaussian reference law $\mathcal N(\bar d_t,q_t)$ on $K_t$. This Bayes factorization uses the signal path for the Gaussian likelihood kernel, the autonomous variance recursion for the uncensored likelihood coordinate $\tilde v_t$, and review outcomes for support restrictions. Continuous public signal updating of the dated file is closed by the smoother block
\[
  \mathcal H_t^\tau=(\tilde m_t,\widehat x_{\tau\mid t},P_{\tau\mid t},C_{\tau t}),
\]
where $\tilde m_t$ is the uncensored signal likelihood mean, $\widehat x_{\tau\mid t}=\E[X_\tau\mid Y_{[0,t]}]$, $P_{\tau\mid t}=\Var(X_\tau\mid Y_{[0,t]})$, and $C_{\tau t}=\Cov(X_\tau,X_t\mid Y_{[0,t]})$ in the reference Gaussian system. The public mean jump is computed by conditioning the pre-jump reference law, or equivalently the uncentered smoother law, on the review event. After that correction the announced gap is recentered, so its post-jump public mean is zero. On a path the announced public state is therefore
\[
  S_t=(m_t,\tilde v_t,\mathcal H_t^\tau,R_t,C_t,A_t,\bar d_t,q_t,K_t),
\]
where $R_t=(R_t^j)_{j\in P\setminus\{0\}}$ is the reputation vector and $A_t=t-\tau$ is the public age of the record. To avoid overloading notation below, a \emph{full review state} includes the public mean and smoother block and is denoted by
\[
  \widehat\xi=(m,\mathcal H,\xi),
  \qquad
  \xi=(\tilde v,R,c,a,\bar d,q,K),
\]
where $\xi$ is the translation-invariant review slice used by interval policy maps. Transition operators use $\widehat\xi$; retention intervals and monotonicity statements use the quotient slice $\xi$. If the episode is resolved, set $C_t=\varnothing$, $A_t=\bar d_t=q_t=0$, $K_t=\{0\}$, and by convention $d_t^{\varnothing}:=0$. In Section~\ref{sec:dynamic}, I show that $(S_t,d_t^{C_t})$ is the exact recursive pair. In the race, the unresolved state uses the same common hidden gap together with public branch-status indicators.

\subsection{Common custody-region audit separation and the single-kernel filter}\label{subsec:audit_separation}

The audit and custody records are public objects with different likelihoods. At a review date, let $o\in\{\mathrm{ret},\mathrm{relay},\mathrm{term}\}$ denote the custody outcome and let $s$ denote the audit mark. The custody outcome determines which file states remain possible. The audit mark moves the reputation state.

\begin{assumption}[Common custody-region audit separation]\label{assumption:audit_separation}
Fix a public review slice $\xi$ and a latent audit type $\theta\in\Theta_i$. Conditional on $\xi$, the hidden gap $d$ has reference density $\phi(d;\bar d,q)$ on support $K$. The custody outcome is generated by a custody region $B_i(\xi)$ common across audit types:
\[
  o=\mathrm{ret}\quad\Longleftrightarrow\quad d\in B_i(\xi),
\]
with relay or terminal disclosure on the complement, as appropriate. Conditional on $(o,\xi,\theta)$, the audit mark $s$ is generated by a public audit experiment with likelihood $\ell_\theta(s\mid o,\xi)$ and is independent of $d$. Thus latent audit quality enters the audit experiment while the custody region is pinned down by the public review slice and the realized file gap.
\end{assumption}

Institutionally, the common custody region is a public case-handling threshold; latent audit quality is learned from the subsequent audit mark. The type-indexed custody variant, in which Bayes' rule yields a finite weighted support rather than a single truncated Gaussian kernel, is treated in the supplementary online appendix (Online Appendix OB).

\begin{proposition}[Common custody-region Bayes update]\label{prop:audit_separation}
Under Assumption~\ref{assumption:audit_separation}, the posterior after a retention event and audit mark $s$ factorizes as
\[
  p(d,\theta\mid \mathrm{ret},s,\xi)
  =
  p(d\mid \mathrm{ret},\xi)\,p(\theta\mid s,\mathrm{ret},\xi),
\]
where
\[
  p(d\mid \mathrm{ret},\xi)
  =
  \frac{\phi(d;\bar d,q)\mathbf 1\{d\in K\cap B_i(\xi)\}}
       {\int_{K\cap B_i(\xi)}\phi(x;\bar d,q)\,dx}.
\]
For a binary audit type with prior $\pi=\Prob(\theta=H\mid\xi)$,
\[
  \pi^+(s,o,\xi)
  =
  \frac{\pi\ell_H(s\mid o,\xi)}
       {\pi\ell_H(s\mid o,\xi)+(1-\pi)\ell_L(s\mid o,\xi)}.
\]
The same statement holds for relay or terminal disclosure after replacing $K\cap B_i(\xi)$ by the corresponding custody event set.
\end{proposition}

\subsection{Timing and equilibrium concept}\label{subsec:model_timing}

Between public events, the fundamental and public signal evolve continuously. In the baseline protocol, reputation changes through the audit transition maps at public review outcomes. At a public review mark, a nonterminal custodian either relays the sealed record to the next node on the route or retains it. The final custodian either terminally discloses the record to the decision maker or retains it. In the race, terminal disclosure by either branch resolves the episode.

Under quadratic loss the DM chooses $y_t=m_t$. All the statements in the main text refer to pure Markov perfect Bayesian equilibria of the institutional model. A pure strategy specifies a review action as a measurable function of the current public state and, for the current holder, the realized gap. In the next subsection, I fix the interval-policy assessments and verification conditions used for the main results. The reputation transitions used there are public audit state updates motivated by the audit experiment in Appendix~\ref{app:extensions}. The latent type foundation is summarized by those public audit state transitions, and custody regions are pinned down by the public state in the single-kernel benchmark.

\subsection{Interval policy assessments and MPBE}\label{subsec:interval_policy_mpbe}

In this subsection, I collect the equilibrium objects used throughout Sections~\ref{sec:dynamic} and~\ref{sec:network}. An admissible support is a finite interval union
\[
  K=\bigcup_{\ell=1}^{L}[\underline k_\ell,\overline k_\ell],
\]
with ordered components and with endpoints in $\overline{\R}=\R\cup\{-\infty,\infty\}$ when exterior rays are needed. The public belief over the current hidden gap is the Gaussian reference kernel $\mathcal N(\bar d,q)$ conditioned on this support. Throughout, $q$ is the variance of the untruncated reference kernel; the conditional variance after censoring is the corresponding truncated moment.

\begin{definition}[Interval policy assessment]\label{def:interval_policy_assessment}
An interval-policy assessment consists of the following objects.
\begin{enumerate}[label=(\roman*),nosep]
  \item For each feasible full review state $\widehat\xi$, with quotient slice $\xi$ and current custodian $c=i$, sender $i$ is assigned an interval $I_i(\xi)\subseteq\R$, possibly empty. Here $\widehat\xi=(m,\mathcal H,\xi)$ and $\xi=(\tilde v,R,c,a,\bar d,q,K)$. The review date retention and forward sets are
  \[
    W_i(\xi)=K(\xi)\cap I_i(\xi),
    \qquad
    F_i(\xi)=K(\xi)\setminus W_i(\xi).
  \]
  Retention is prescribed on $W_i(\xi)$, with weak review-action ties assigned to $F_i(\xi)$; finite relative-frontier ties have zero probability under the continuous gap law. Relay, or terminal disclosure if the next node is the decision maker, is prescribed on $F_i(\xi)$.
  \item Beliefs are updated by Bayes' rule after every on-path review outcome, using the censored Gaussian update generated by the observed event and the support recursion in Section~\ref{sec:dynamic}.
  \item The relative frontier of the retention set is
  \[
    \partial_K W_i(\xi)
    =
    \overline{W_i(\xi)}^{\,K}
    \cap
    \overline{K(\xi)\setminus W_i(\xi)}^{\,K},
  \]
  where closure is relative to the order topology of $K(\xi)$. Hence the relevant frontier is the policy cut inside the feasible support, selected independently of the component endpoints of $K(\xi)$.
  \item For each zero-probability review outcome $o\in\{\mathrm{ret},\mathrm{relay},\mathrm{term}\}$, the assessment specifies an ordered action cut $\zeta_i^o(\xi)=(e_i^o(\xi),s_i^o(\xi))$. If $\partial_K W_i(\xi)$ is nonempty, $e_i^o(\xi)$ is a relative frontier point. If it is empty, $e_i^o(\xi)$ is a specified endpoint of $K(\xi)$. The side $s_i^o(\xi)$ selects the outcome side of the cut: the retained side for retention and the forward side for relay or terminal disclosure. The off-path posterior is the normalized Gaussian reference kernel on the selected side $C_i^o(\zeta,\xi)\subseteq K(\xi)$; whenthis side collapses to a frontier point, the posterior is the corresponding one-sided degenerate weak limit. Lemma~\ref{lem:frontier_completion} shows that this completion is the unique limit of admissible interval trembles approaching the ordered cut.
\end{enumerate}

\end{definition}

The ordered-cut completion is a belief convention only at zero-probability review actions. On path, beliefs are the exact censored-Gaussian Bayes updates generated by the realized custody event.

The verification step uses the no-review generator $\mathcal A^Z$ of the exact joint state $Z=(S,d)$ and the review date operators $\mathcal M^i$ and $\mathcal K^i$ defined in Section~\ref{sec:dynamic}. The operator $\mathcal M^i$ is relay or terminal disclosure, according to the current position of $i$ in the route, and $\mathcal K^i$ is retention.

\begin{proposition}[Sufficient verification of interval-policy assessments]\label{prop:interval_policy_mpbe_verification}
Fix now an interval-policy assessment. Suppose that for each sender $i$ there is a locally bounded continuation value $V^i$ that is measurable in the exact state, satisfies the transversallity condition for discounted payoffs, and solves
\begin{equation}
\begin{aligned}
  \rho V^i(z)=u_i(z)+\mathcal A^Z V^i(z)
  &+\mathbf 1\{C(z)=i\}\lambda^i(S)
    \Bigl(\max\{\mathcal M^iV^i(z),\mathcal K^iV^i(z)\}-V^i(z)\Bigr)\\
  &+\sum_{j\neq i}\mathbf 1\{C(z)=j\}\lambda^j(S)
    \Bigl(\mathcal T_j^\sigma V^i(z)-V^i(z)\Bigr),
\end{aligned}
  \label{eq:model_mpbe_bellman}
\end{equation}
where $\mathcal T_j^\sigma$ is the public jump operator generated by sender $j$'s equilibrium review action. The active choice term for $i$ applies only when $i$ is current custodian. The passive review terms are $0$ on $i$'s own custody region and are relevant off custody only when another sender is active. On $i$'s retention region the equation is
\begin{equation}
  \rho V^i(z)=u_i(z)+\mathcal A^Z V^i(z)
  +\lambda^i(S)\bigl(\mathcal K^iV^i(z)-V^i(z)\bigr).
  \label{eq:model_mpbe_restricted}
\end{equation}
At every feasible review state the one-step inequalities are
\[
  \mathcal K^iV^i(z)\ge \mathcal M^iV^i(z)
  \quad\text{on } W_i(\xi),
  \qquad
  \mathcal M^iV^i(z)\ge \mathcal K^iV^i(z)
  \quad\text{on } F_i(\xi),
\]
where we have value matching at every finite relative frontier point. Public beliefs are Bayes-consistent on-path and are completed off-path by the ordered-cut rule in Definition~\ref{def:interval_policy_assessment}. Then the assessment is a pure Markov perfect Bayesian equilibrium of the public custody review game on its admissible state space.
\end{proposition}

\subsection{Benchmark protocol and extensions}\label{subsec:model_scope}

Public docketing, custody logs, and publicly observed review dates generate the censoring recursion in Proposition~\ref{prop:state_compression}. The single operative record and no-supersession protocol keep that recursion exact. Theorem~\ref{thm:dynamic_unraveling} adds recurrent public review, maintained compact-slice discipline, and a primitive age envelope that makes one more skipped active review too costly once the file is old enough. The main theorems remain within this benchmark; bounded-memory variants and related institutional protocols are collected in the supplementary online appendix.


\section{Static Strategic Disclosure in Networks}\label{sec:static}

This section records the static hard-information benchmark that motivates public review. Direct disclosure has the familiar unraveling force. When the path contains an intermediary whose bias points in the opposite direction, anonymous blocking creates dark states. Logged public review turns that static obstruction into endogenous delay: custody is visible, review dates arrive publicly, and retention itself becomes the signal. Reputational payofs and clocks are suppressed here.

Consider first one expert and the decision maker. A scalar state $\theta\in\Theta=[\underline \theta,\bar \theta]\subseteq\R$ is drawn from an atomless prior $F$ with full support. Only the expert observes $\theta$. The DM chooses $y\in\R$ and has payoff $u_0(y,\theta)=-(y-\theta)^2$, while the expert has payoff $u_e(y,\theta)=-\alpha_e(y-\theta-b_e)^2$, with $\alpha_e>0$. The expert can disclose the true state or remain silent; messages are verifiable and costless.

\begin{proposition}[Direct unraveling]\label{prop:direct_unraveling}
In any perfect Bayesian equilibrium of the direct expert-to-DM game with verifiable messages, the DM's action equals the realized state almost surely: $y(\theta)=\theta$ for $F$-almost every $\theta$. Equivalently, no positive measure set of states remains permanently undisclosed.
\end{proposition}

Now consider a static expert-to-DM path $P=(e,i_1,\dots,i_k,0)$ on which at most one verifiable message can travel. Each sender $r\in\{e,i_1,\dots,i_k\}$ has payoff $u_r(y,\theta)=-\alpha_r(y-\theta-b_r)^2$, with publicly known bias $b_r$ and $\alpha_r>0$.

\begin{definition}[Bias aligned and bias reversing paths]\label{def:bias_alignment}
A path is \emph{bias aligned} if all biases along it have the same weak sign. It is \emph{bias reversing} if it contains agents with opposite signed biases.
\end{definition}

\begin{proposition}[Static path and tree benchmark]\label{prop:static_path}\label{prop:static_network}
Consider the static disclosure game along a fixed path with verifiable messages.
\begin{enumerate}[label=(\roman*),nosep]
  \item If the path is bias aligned, there exists a PBE with full disclosure: the true state reaches the DM for every realization and $y(\theta)=\theta$.
  \item If the path is bias reversing and blocking is anonymous, so that the DM observes only that no message arrived while the blocker identity is hidden, full truthful transmission on all states cannot be sustained.
\end{enumerate}
Consequently, on the unique expert-to-DM path of a directed tree, full disclosure obtains under bias alignment, whereas anonymous bias reversal prevents full truthful transmission on the path. This benchmark fixes the reference point; the dynamic results below study disclosure speed and eventual transmission in logged review institutions.
\end{proposition}

\section{Dynamic Disclosure Through Public Review}\label{sec:dynamic}

We analyze disclosure along a fixed expert-to-DM path
\[
  P=(e,i_1,\dots,i_k,0).
\]
Along such a path, public retention censors the holder's latent gap and public relay can create support holes while preserving an exact scalar likelihood representation. Interval policies add finite endpoint closure. The section proceeds from the exact recursion to the compact interval-policy benchmark and then to old-record resolution.

\subsection{State recursion on a path}\label{subsec:dynamic_beliefs}

Fix a path episode with publicly docketed operative file $\varrho_\tau=(X_\tau,\tau)$. When the current public custodian is $C_t=i\neq\varnothing$, intermediary $i$ privately knows the sealed content and hence
\[
  \mu_t^i = \E[X_t\mid X_\tau,\mathcal F_t^p],
  \qquad
  d_t^i := \mu_t^i-m_t.
\]
Public docketing makes age and custody public; the hidden object is the holder's scalar informational advantage $d_t^i$.

For any Borel $B\subseteq\R$, let
\[
  \mathfrak m(B;\bar d,q):=\E[Z\mid Z\in B],
  \qquad
  \mathfrak q(B;\bar d,q):=\Var(Z\mid Z\in B),
\]
for $Z\sim\mathcal N(\bar d,q)$. These operators recover exact censored moments from the announced Gaussian reference kernel; In Appendix~\ref{app:diffusion_prelim}, I give the interval union formulas.

\begin{lemma}[Gaussian kernel-support factorization]\label{lem:kernel_support_factorization}
Fix a public history before terminal disclosure. If the public signal path gives the reference gap density $\phi(d;\bar d,q)$ and the docket history leaves feasible support $K$, then
\[
  p(d\mid \text{public signals},\text{ docket events})
  =
  \frac{\phi(d;\bar d,q)\mathbf 1\{d\in K\}}{\int_K\phi(x;\bar d,q)\,dx}.
\]
If a review event replaces $K$ by $H\subseteq K$, then the public mean jump is $\Delta m=\mathfrak m(H;\bar d,q)$ and the announced post-event kernel-support pair is $(\bar d-\Delta m,q,H-\Delta m)$. Thus censoring changes support and truncated moments, while the reference variance coordinate $q$ is unchanged at the censoring date.
\end{lemma}

\begin{proposition}[Exact filtering for the latent gap]\label{prop:state_compression}
Suppose the fundamental is Ornstein-Uhlenbeck, the public signal is linear Gaussian, the operative file's original timestamp and chain of custody are publicly logged, and the benchmark episode contains a single pending sealed file. Fix any profile of pure review date policies on a simple path. Then conditional on public history there exist publicly measurable $(\bar d_t,q_t,K_t)$ such that
\[
  d_t^{C_t}\mid \mathcal F_t^p
  \sim
  \mathcal N(\bar d_t,q_t)\ \text{conditioned on}\ d_t^{C_t}\in K_t,
\]
where $K_t\subseteq\R$ is the feasible support generated by the realized public history. The pair $(\bar d_t,q_t)$ is the Gaussian likelihood kernel for the latent gap. The exact public posterior is that kernel conditioned on $K_t$, combining the public signal path with review censoring events. At docketing, $\bar d_\tau=0$, $q_\tau=\tilde v_\tau$, and $K_\tau=\R$.

If no public review event or relay occurs on $[s,t]$, there exist publicly measurable $\chi_{s,t}>0$ and $\psi_{s,t}\in\R$ such that
\[
  d_t^{C_t}=\chi_{s,t}d_s^{C_s}+\psi_{s,t},
  \qquad
  K_t=\chi_{s,t}K_s+\psi_{s,t}.
\]
Thus support transport is exact. The reference coordinates $(\bar d_t,q_t)$ are recomputed from the realized public signal path on $[s,t]$ through the smoother block, while the feasible support follows the displayed affine transport.

At a public review date with current public slice $\xi$, let $B_i(\xi)\subseteq K_{t-}$ be the current retention set. Observed retention and relay update the announced reference stateby
\[
  K_t^{\mathrm{ret}} = \bigl(K_{t-}\cap B_i(\xi)\bigr)-\Delta m_t^{\mathrm{ret}},
  \qquad
  \bar d_t^{\mathrm{ret}} = \bar d_{t-}-\Delta m_t^{\mathrm{ret}},
  \qquad
  q_t^{\mathrm{ret}}=q_{t-},
\]
\[
  K_t^{\mathrm{relay}} = \bigl(K_{t-}\cap B_i(\xi)^c\bigr)-\Delta m_t^{\mathrm{relay}},
  \qquad
  \bar d_t^{\mathrm{relay}} = \bar d_{t-}-\Delta m_t^{\mathrm{relay}},
  \qquad
  q_t^{\mathrm{relay}}=q_{t-},
\]
with
\[
  \Delta m_t^{\mathrm{ret}}=\mathfrak m\bigl(K_{t-}\cap B_i(\xi);\bar d_{t-},q_{t-}\bigr),
  \qquad
  \Delta m_t^{\mathrm{relay}}=\mathfrak m\bigl(K_{t-}\cap B_i(\xi)^c;\bar d_{t-},q_{t-}\bigr).
\]
The equalities $q_t^{\mathrm{ret}}=q_t^{\mathrm{relay}}=q_{t-}$ refer to the reference-kernel variance. The corresponding exact post-review gap variances are
\[
  \mathfrak q\bigl(K_{t-}\cap B_i(\xi);\bar d_{t-},q_{t-}\bigr)
  \qquad\text{and}\qquad
  \mathfrak q\bigl(K_{t-}\cap B_i(\xi)^c;\bar d_{t-},q_{t-}\bigr).
\]
At relay the sealed content is unchanged, so the downstream custodian inherits the same file-based posterior mean and therefore $d_t^j=d_{t-}^i-\Delta m_t^{\mathrm{relay}}$.
\end{proposition}

Appendix~\ref{app:diffusion_prelim} records the closed smoother equations generating the reference coordinates $(\bar d_t,q_t)$ and the public affine support transport.

\begin{corollary}[Interval union closure on a path]\label{cor:endpoint_union_closure}
If, on each feasible public slice, every review date retention event set is generated by a finite interval-union policy, then $K_t$ remains a finite union of intervals along the path. Consequently, on a fixed path the continuation problem has a finite endpoint representation after encoding the ordered endpoints of those components together with
\[
  S_t=(m_t,\tilde v_t,\mathcal H_t^\tau,R_t,C_t,A_t,\bar d_t,q_t,K_t).
\]
\end{corollary}

\begin{corollary}[Linear component bound under interval policies]\label{cor:endpoint_component_bound}
If, in addition, every review date policy interval $I_i(\xi)$ is a single interval, so the actual retention event is $K(\xi)\cap I_i(\xi)$, then after $r_t$ public relays the support $K_t$ has at most $1+r_t$ connected components; in particular, on a path with $k$ downstream custody transfers the component count is at most $k+1$. A relay intersects $K$ with the complement of one interval and can split only the component it intersects.
\end{corollary}

\begin{example}[Support holes and exact censoring]\label{ex:support_hole_main}
Let $d\sim\mathcal N(0,1)$. Suppose one retention restricts support to $(-.6,.6)$ and a later relay removes $(-.2,.2)$. The exact support is $K=(-.6,-.2]\cup[.2,.6)$, with mean zero and variance
\[
  \frac{2\int_{.2}^{.6} z^2\phi(z)\,dz}{2[\Phi(.6)-\Phi(.2)]}\approx .169.
\]
A Gaussian shortcut matching these two moments assigns positive mass to $(-.15,.15)$; the exact recursion assigns zero. Endpoint support is therefore payoff-relevant: retaining the support hole changes the feasible waiting states and can change the next review action. In Figure~\ref{fig:support_censoring} below, I illustrate the support operation.
\end{example}

\begin{figure}[t]
\centering
\begin{tikzpicture}[x=4.7cm,y=.75cm]
  \node[anchor=east] at (-.03,2) {initial support $K$};
  \draw[very thick] (.05,2) -- (.95,2);
  \node[anchor=east] at (-.03,1) {retention $K\cap I$};
  \draw[very thick] (.25,1) -- (.72,1);
  \node[anchor=east] at (-.03,0) {relay $K\cap I^c$};
  \draw[very thick] (.05,0) -- (.25,0);
  \draw[very thick] (.72,0) -- (.95,0);
  \draw[dashed] (.25,-.22) -- (.72,-.22);
  \node[anchor=west] at (.74,-.22) {removed interval};
  \draw[thin] (.25,2.2) -- (.25,-.35);
  \draw[thin] (.72,2.2) -- (.72,-.35);
\end{tikzpicture}
\caption{Public review changes beliefs by changing support. A retention policy interval $I$ intersects the feasible support $K$; a relay event applies the complementary restriction and can leave a support hole that a moment-matched Gaussian shortcut would fill.}
\label{fig:support_censoring}
\end{figure}

\paragraph{Docketing contrast.}
If custody logs and review marks are suppressed, the same nonterminal silence is not a review action. The public cannot tell whether the file was unreviewed, reviewed and retained, or sitting with another custodian, hence Bayes' rule averages over latent custodians, dates, and retained sets. The posterior is a mixture over histories rather than a Gaussian kernel on a finite endpoint support. Public docketing therefore does not relabel delay; it creates the observed event that makes the support recursion exact.

\begin{proposition}[Exact recursion for the announced state]\label{prop:full_state_recursion}
Under Assumption~\ref{assumption:markov_rep} and Corollary~\ref{cor:endpoint_union_closure}, the announced public state
\[
  S_t=(m_t,\tilde v_t,\mathcal H_t^\tau,R_t,C_t,A_t,\bar d_t,q_t,K_t)
\]
is Markov on the single-file path benchmark. If no public review event occurs on $[s,t]$ while $C_u=i$, then there is a Borel functional of the time-$s$ announced state and the realized public signal segment $Y_{[s,t]}$ such that
\[
  \Phi^{i}_{s,t}:\bigl((m,\tilde v,\mathcal H,R,i,a,\bar d,q,K),Y_{[s,t]}\bigr)
  \mapsto
  \bigl(m',\tilde v',\mathcal H',R,i,a+t-s,\bar d',q',\chi_{s,t}K+\psi_{s,t}\bigr),
\]
where $\tilde v'$ and $\mathcal H'$ are generated by the reference filter and smoother, $(\bar d',q')$ is the induced likelihood kernel for the hidden gap, and $m'$ is tracked recursively by the smoother and prior review jumps. At a review, the public mean correction is obtained by applying the censored moment operator to the uncentered gap law just before the event; the announced gap is then recentered. Reputation remains unchanged between review outcomes.

At a public review date of current custodian $i$ with full review state $\widehat\xi=(m,\mathcal H,\xi)$, quotient slice $\xi=(\tilde v,R,c=i,a,\bar d,q,K)$, and retention set $B_i(\xi)$, the announced state jumps according to
\[
T_i^{\mathrm{ret}}(\widehat\xi)=
\begin{aligned}[t]
  \bigl(&m+\Delta m_i^{\mathrm{ret}}(\xi),\,\tilde v,\,\mathcal H,\,R_i^{\mathrm{ret}}(\xi),\, i,\, a,\\
  &\bar d-\Delta m_i^{\mathrm{ret}}(\xi),\, q,\, (K\cap B_i(\xi))-\Delta m_i^{\mathrm{ret}}(\xi)\bigr),
\end{aligned}
\]
\[
T_{i\to j}^{\mathrm{relay}}(\widehat\xi)=
\begin{aligned}[t]
  \bigl(&m+\Delta m_i^{\mathrm{relay}}(\xi),\,\tilde v,\,\mathcal H,\,R_{i\to j}^{\mathrm{relay}}(\xi),\, j,\, a,\\
  &\bar d-\Delta m_i^{\mathrm{relay}}(\xi),\, q,\, (K\cap B_i(\xi)^c)-\Delta m_i^{\mathrm{relay}}(\xi)\bigr),
\end{aligned}
\]
and terminal disclosure sends the state to
\[
  T_i^{\mathrm{term}}(\widehat\xi,d)=\bigl(m+d,\,\tilde v,\,\mathcal H,\,R_i^{\mathrm{term}}(\xi,d),\,\varnothing,\,0,\,0,\,0,\,\{0\}\bigr).
\]

The maps $R_i^{\mathrm{ret}}$, $R_{i\to j}^{\mathrm{relay}}$, and $R_i^{\mathrm{term}}$ are the public audit transitions associated with the observed review outcome. In the main text, they move the reputation state and do not enter the sealed evidence filter: evidence moments after a review are recovered by applying $\mathfrak m$ and $\mathfrak q$ to the retained or relayed truncation set. Appendix~\ref{app:extensions} gives a foundation with latent audit quality, in which the same transitions are Bayesian posteriors over that quality.

\end{proposition}

Now write $Z_t=(S_t,d_t^{C_t})$. For each intermediary $i$, let $\mathcal A^Z$ denote the no-review generator of the full exact joint state $Z_t$. It includes the drift, diffusion, and cross-variation terms of the public signal likelihood coordinates, the smoother derived kernel, the support transport, the exact censored public mean, and the current holder's latent gap. Let $\mathcal M^i$ be the review date disclosure operator, let $\mathcal K^i$ be the review date retention operator, and let $\mathcal T_j^\sigma$ be the public jump operator induced by sender $j$'s equilibrium review action. With review intensity $\lambda^j(S)$ for each public custodian $j$, the sender $i$'s Bellman equation can then be written
\begin{equation}
\begin{aligned}
  \rho V^i(z)=u_i(z)+\mathcal A^Z V^i(z)
  &+\mathbf 1\{C(z)=i\}\lambda^i(S)
    \Bigl(\max\{\mathcal M^iV^i(z),\mathcal K^iV^i(z)\}-V^i(z)\Bigr)\\
  &+\sum_{j\neq i}\mathbf 1\{C(z)=j\}\lambda^j(S)
    \Bigl(\mathcal T_j^\sigma V^i(z)-V^i(z)\Bigr),
\end{aligned}
  \label{eq:hjb_main}
\end{equation}
where the active choice term for $i$ is present only on $i$'s custody region. The passive review term is $0$ there and is relevant off custody only when another sender is active.

Assumption~\ref{assumption:markov_rep} isolates the main single-file path benchmark. Interval restrictions enter the equilibrium class studied next.

\begin{assumption}[Benchmark review environment]\label{assumption:markov_rep}
Along the path $P=(e,i_1,\dots,i_k,0)$:
\begin{enumerate}[label=(\roman*),nosep]
  \item the fundamental follows an Ornstein-Uhlenbeck diffusion and the public signal is linear Gaussian;
  \item the benchmark episode contains one publicly docketed sealed file and no supersession before that file is resolved;
  \item review calendars and realized review dates are public and independent of privately held record content;
  \item reputational review transitions are public audit state transitions and are conditionally separated from the scalar evidence filter;
  \item the DM chooses $y_t=m_t$ at each date.
\end{enumerate}
\end{assumption}

\begin{remark}[Off-path reviews and degenerate waiting sets]\label{rem:off_path_review_beliefs}

Beliefs after review actions assigned probability zero are completed by the ordered cut specified in Definition~\ref{def:interval_policy_assessment}. If such a retention, relay, or terminal disclosure is observed, the posterior is the weak limit of the corresponding censored Gaussian update as interval policies approach $\zeta_i^o(\xi)$ from the side selected by that action. Lemma~\ref{lem:frontier_completion} shows that, once the endpoint and side are fixed, the limit does not depend on the particular trembles used to approach it.

\end{remark}

\subsection{Compact interval-policy benchmark}\label{subsec:compact_benchmark}

The equilibrium result is a benchmark verification theorem. The exact filter is Bayesian; the affine Ornstein-Uhlenbeck subclasses summarized below give the primitive single-crossing geometry behind interval review responses.

\begin{assumption}[Compact OU-quadratic interval benchmark]\label{assumption:compact_interval_benchmark}
On the finite path $P=(e,i_1,\dots,i_k,0)$, feasible unresolved slices belong to a compact domain closed under the support transport and review date censoring maps in Proposition~\ref{prop:full_state_recursion}. Material payoffs are quadratic in the DM's action, the fundamental and signal form an Ornstein-Uhlenbeck linear Gaussian system, and each review policy is a single interval in the current feasible support. At each review boundary, the gap maps for relay and for continuation after retention agree at endpoints and preserve the order on the slice selected by the relevant side. The active review gain has strict sublevel sets that are monotone in each coordinate when approached from that side, and cannot bypass any component of the feasible support, which is a finite union of intervals. Audit recertification has weight $\chi_i\in(0,1)$ and keeps reputation in a compact interior region where the discipline margin used below applies whenever an active review is skipped.

\end{assumption}

\begin{theorem}[Compact OU-quadratic interval-policy verification]\label{thm:primitive_ou_quadratic_audit}
Under Assumptions~\ref{assumption:markov_rep} and~\ref{assumption:compact_interval_benchmark}, the compact OU-quadratic conditions verify a pure interval-policy MPBE of the single-file path game on the benchmark domain. Conditional on each public history, the hidden holder gap has the exact Gaussian-likelihood representation of Proposition~\ref{prop:state_compression}; under the interval-policy profile its feasible support is the finite interval union in Corollary~\ref{cor:endpoint_union_closure}. At every feasible review slice $\xi$ for current custodian $i$, equilibrium retention is
\[
  W_i(\xi)=K(\xi)\cap I_i^\ast(\xi)
\]
for a slice-specific interval $I_i^\ast(\xi)$; outside $W_i(\xi)$ the file moves forward by relay or terminal disclosure.
\end{theorem}

\begin{proposition}[Disclosure ladder on the compact benchmark]\label{prop:path_ladder}
On the benchmark class of Theorem~\ref{thm:primitive_ou_quadratic_audit}, public review generates a finite disclosure ladder. Each active review either keeps the file inside the current interval $K(\xi)\cap I_i^\ast(\xi)$ or moves it downstream. Each move updates the exact endpoint state by the censoring recursion, and terminal-stage forward movement is terminal disclosure.
\end{proposition}

A concrete affine OU subclass shows that interval policies are nonempty. At the terminal holder's review, let the retention gain and radius be
\[
G_i(d,a):=C_i e^{-\lambda a}-\eta_i(d-p_i)^2-\beta_i\underline\delta_i,
\qquad
r_i(a):=\left[\frac{C_i e^{-\lambda a}-\beta_i\underline\delta_i}{\eta_i}\right]_+^{1/2}.
\]
Then retention is exactly \(K\cap[p_i-r_i(a),p_i+r_i(a)]\). On compact slices away from the boundary, affine support transport, affine maps for relay and for the proxy after retention, and signed concave continuation increments imply the endpoint inequalities and preclude bypassing support components. Hence Assumption~\ref{assumption:compact_interval_benchmark} holds (details deferred to Online Appendix OB).

\subsection{Discipline-assisted old-record resolution}\label{subsec:dynamic_equilibrium}

The next assumptions impose recurrence, audit-reputation discipline, and the age envelope for resolution in the single-file review model. I include the audit experiment foundation, audit refresh, and the dated-record age envelope in Appendix~\ref{app:extensions}. 

\begin{assumption}[Review recurrence]\label{assumption:LQG_path}
For each sender $i\in P\setminus\{0\}$ and every history on which the operative file reaches $i$ and remains outstanding after time $t$, let $\widetilde\lambda_u^i$ be the predictable review intensity evaluated along the unresolved continuation path on which no further active review of $i$ occurs between $t$ and $u$. Then
\[
  \int_t^\infty \widetilde\lambda_u^i\,du = \infty
\]
almost surely on that event.
\end{assumption}

\begin{assumption}[Review transparency discipline]\label{assumption:uniform_discipline}
For each sender $i\in P\setminus\{0\}$ there exist audit types $H$ and $L$, a compact unresolved reputation region relevant along the path, and a constant $\underline\delta_i>0$ such that whenever $i$ currently holds the operative file, the current public state lies in that region, and a public active review is skipped by retention, the expected unweighed continuation contribution generated by the reputational term $g_i(R_t^i)$ falls by at least $\underline\delta_i$ relative to immediate relay or terminal disclosure.
\end{assumption}

Under the audit experiment foundation that I present in Appendix~C, public recertification keeps audit beliefs in an interior region and prompt resolution generates a more informative Bayes-plausible audit experiment than skipped retention. The following primitive audit refresh condition gives an invariant reputation region and the continuation-value margin used by Assumption~\ref{assumption:uniform_discipline}.

\begin{corollary}[Audit refresh verifies interior discipline]\label{cor:audit_refresh_discipline}
Suppose the audit state is a posterior $\pi_t^i\in(0,1)$ over audit types. Before each active review, public recertification mixes the posterior with an interior benchmark $\pi_i^0\in(0,1)$ with weight $\chi_i\in(0,1)$. Let $G_i(\pi)$ be the unweighted discounted continuation contribution of the audit-reputation state after posterior $\pi$; in a one-shot or absorbing post-review benchmark this is a persistence/discounting multiple of $g_i(\rho_i(\pi))$. Suppose $G_i$ is twice continuously differentiable on the recertified interior interval with $G_i''\ge\gamma_i>0$, and realized prompt and skipped-review posteriors remain supported there. If prompt review is a uniformly more informative Bayes-plausible experiment, with both posterior experiments having mean $\pi$ and variance gap at least $\eta_i>0$, then unresolved audit posteriors remain in a compact interval $[\underline\pi_i,\overline\pi_i]\subset(0,1)$ and $\underline\delta_i=\gamma_i\eta_i/2>0$ verifies Assumption~\ref{assumption:uniform_discipline}.
\end{corollary}

At a skipped active review, retention must earn material gain at least $\beta_i\underline\delta_i$ to ofset the reputational loss from not moving the file forward. Reputation therefore shrinks the retention set, and the age envelope makes this shrinkage decisive for old files.

For an operative file $\varrho$ currently held by sender $i$, let $G_i^{\mathrm{mat}}(d,\xi;\varrho)$ denote the continuation value material gain from one more retention at a review date, evaluated at realized gap $d$ and quotient slice $\xi$, after shutting down the reputational term. This object is measurable in the exact announced state and the operative file's date.

\begin{assumption}[Compact-slice age envelope for material retention gains]\label{assumption:material_age_envelope}
For each sender $i\in P\setminus\{0\}$ there exist a deterministic nonincreasing, right-continuous function $\overline G_i:[0,\infty)\to[0,\infty)$ with $\overline G_i(a)\downarrow0$ such that, on every unresolved equilibrium history whose reputation and gap slice lie in the maintained compact unresolved region of Assumption~\ref{assumption:uniform_discipline},
\[
  G_i^{\mathrm{mat}}(d,\xi;\varrho)\le \overline G_i(a)
\]
for every feasible realized gap $d$ and every feasible quotient slice $\xi$ with public file age $a$.
\end{assumption}

Assumption~\ref{assumption:material_age_envelope} is the primitive age-decay restriction behind deterministic old-record resolution on compact slices. In the dated record Ornstein-Uhlenbeck benchmark, it follows from exponential decay of the file's incremental effect on future public actions, uniformly on compact unresolved slices; see Appendix~\ref{app:extensions}. Define now deterministic review-discipline age
\[
  a_i^\dagger
  :=
  \inf\{a\ge0:\overline G_i(a)\le \beta_i\underline\delta_i\},
\]
which is finite whenever $\beta_i\underline\delta_i>0$. Since $\overline G_i$ is nonincreasing and right-continuous, the sublevel set contains its infimum, so $a_i^\dagger$ is a minimum and $\overline G_i(a_i^\dagger)\le \beta_i\underline\delta_i$.

\begin{theorem}[Discipline-assisted compact-slice old-record resolution]\label{thm:dynamic_unraveling}
Consider a pure interval-policy MPBE of the single-file path game on $P=(e,i_1,\dots,i_k,0)$ under Assumptions~\ref{assumption:markov_rep}, \ref{assumption:LQG_path}, \ref{assumption:uniform_discipline}, and \ref{assumption:material_age_envelope}. The deterministic conclusion is conditional on the unresolved history remaining in the maintained compact reputation and feasible-gap slice entering Assumptions~\ref{assumption:uniform_discipline}--\ref{assumption:material_age_envelope}. Let the bias profile $(b_e,b_{i_1},\dots,b_{i_k})$ be arbitrary, including bias reversals.

At any public active review of sender $i$ at which the operative file is still outstanding, the reputation state and feasible gap slice are in the maintained compact region, and the public age of the file satisfies $a\ge a_i^\dagger$, forward movement is weakly optimal and, under the tie convention in Definition~\ref{def:interval_policy_assessment}, the review induces relay or terminal disclosure. Consequently, conditional on the maintained compact-slice event, if a sender $i$ receives the file and the file remains outstanding at $i$ until age $a_i^\dagger$, the next active public review of $i$ after that age moves the file forward almost surely. Under review recurrence, that review occurs in finite calendar time almost surely; since custody only moves forward on a finite path, terminal disclosure follows in finite calendar time almost surely as well.
\end{theorem}

\begin{corollary}[Unconditional Gaussian-tail old-record bound]\label{cor:gaussian_tail_old_record}
In the same review environment, suppose the hidden gap reference support is unbounded but conditional moments are finite and the material retention gain satisfies an expected age envelope
\[
  \E\bigl[G_i^{\mathrm{mat}}(d,\xi;\varrho)\mid\xi,a\bigr]\le \overline G_i^E(a),
  \qquad \overline G_i^E(a)\downarrow0.
\]
Then any active review after $a_i^E:=\inf\{a:\overline G_i^E(a)\le\beta_i\underline\delta_i\}$ has nonpositive expected incremental value from retention. If $\Prob(G_i^{\mathrm{mat}}(d,\xi;\varrho)>\beta_i\underline\delta_i\mid\xi,a)\le\varepsilon_i(a)$ with $\varepsilon_i(a)\downarrow0$, the probability that retention is pointwise profitable at an active review after age $a$ is at most $\varepsilon_i(a)$.
\end{corollary}

The theorem applies at review dates. Between review marks, the continuous gap may move across state-dependent regions; the compact-slice age envelope controls the feasible unresolved states to which the deterministic threshold is applied. With unbounded Gaussian support, the same force yields expected and high-probability versions.

\begin{remark}[Selection and robust objects]\label{rem:selection_robust_objects}
The exact filter is the Bayesian recursion for the public state, conditional on the public history and the strategy profile. Section~\ref{sec:network} applies the same support restriction when two certified routes carry the same record. Comparative statics and formation values are evaluated at the verified equilibrium in interval policies for the relevant benchmark; I include a corresponding monotone comparison for the greatest fixed point in the supplementary online appendix (Online Appendix OB).
\end{remark}


\section{Network Design and Formation}\label{sec:network}

Here I study the two-route race from Section~\ref{subsec:model_network}, the minimal overlapping architecture in which two certified branches carry one common record. The first results isolate common-record censoring; the formation results then show how bilateral link consent can miss this value.

\subsection{Common record race and censoring}\label{subsec:network_value}

Fix a feasible network $G$ and let $V_i(G)$ denote agent $i$'s ex ante equilibrium continuation value at the docket date. When $G$ is a line with one expert $e$ and one decision maker $0$, the network game coincides with the single-file path game on the unique route $P_G(e\to 0)$. Writing
\[
  g_\tau:=X_\tau-m_\tau,
\]
we can evaluate the route problem directly under the docket date law of $(S_\tau,g_\tau)$:
\begin{equation}
  V_i(G) = \E\bigl[V^i_{P_G(e\to 0)}(S_\tau,g_\tau)\bigr],
  \label{eq:network_value_decomposition}
\end{equation}
where $V^i_{P_G(e\to 0)}$ is the path continuation value from Proposition~\ref{prop:path_ladder} and Theorem~\ref{thm:dynamic_unraveling}. If the initial condition is already resolved, this is the corresponding no-file continuation value.

In the two-route race, the expert can place certified copies of the same docketed record on both branches. The public observes the common original timestamp, every relay receipt, every public review mark, and which branch first reaches the DM. Because the copies have the same content and date, the branch gaps are identical on every unresolved public slice:
\[
  d_i=d_j=g.
\]
This common gap representation is exact for certified copies of one docketed record. Conditional on the sealed record the race is complete-information, but actions occur only at public review jumps; simultaneous branch reviews have probability zero under the maintained clocks. The scalar common gap is therefore the exact state for the common-record race studied here. I also present extensions treating imperfectly correlated copies as the bivariate Gaussian perturbation of this benchmark in the supplementary online appendix.

\begin{lemma}[Common gap recursion in the race benchmark]\label{lem:race_state}
Consider the single-file review benchmark of Section~\ref{sec:model} and fix now a profile of pure branch policies such that, on each feasible public slice, each branch retains the file at a review date if and only if the common gap lies in a branch interval. Then, conditional on unresolved public history, there exists publicly measurable $(\bar g_t,q_t,K_t,\Xi_t)$ such that the common hidden gap $g_t$ is Gaussian with mean $\bar g_t$ and variance $q_t$ conditioned on $g_t\in K_t$, the branch gaps satisfy $d_i=d_j=g_t$, and $\Xi_t$ records public branch status and route resolution. Consequently, the unresolved two-route race has a finite endpoint public state on the interval-policy class. The scalar common gap is the nonredundant state; stacking the two-branch gaps would produce an exactly singular covariance matrix on every unresolved slice.
\end{lemma}

\begin{corollary}[Interval union closure in the race benchmark]\label{cor:race_union_closure}
Under the hypotheses of Lemma~\ref{lem:race_state} above, if each branch retains at a review date whenever the common gap lies in an interval, then the common feasible support $K_t$ for the hidden gap $g_t$ remains a finite union of intervals along unresolved public histories.
\end{corollary}

\begin{proposition}[Exact recursion for the unresolved state in the race benchmark]\label{prop:race_full_state_recursion}
Under Lemma~\ref{lem:race_state} and Corollary~\ref{cor:race_union_closure}, the unresolved race announced state
\[
  S_t^{\mathrm{TR}}=(m_t,\tilde v_t,\mathcal H_t^\tau,R_t,A_t,\bar g_t,q_t,K_t,\Xi_t)
\]
is Markov. If no public review event occurs on $[s,t]$ while the race remains unresolved, then there is a Borel functional of the time-$s$ unresolved state and the realized public signal segment $Y_{[s,t]}$,
\[
  \Phi^{\mathrm{TR}}_{s,t}:\bigl(S_s^{\mathrm{TR}},Y_{[s,t]}\bigr)\mapsto S_t^{\mathrm{TR}},
\]
whose support component is the exact affine transport $K_t=\chi^{\mathrm{TR}}_{s,t}K_s+\psi^{\mathrm{TR}}_{s,t}$ of the common hidden gap and whose public coordinates update $(m_t,\tilde v_t,\mathcal H_t^\tau,R_t,A_t,\bar g_t,q_t,\Xi_t)$ exactly from the same realized public signal path, elapsed time since docketing, and publicly observed branch status.

At a public review date on branch $b\in\{i,j\}$ with unresolved slice $\xi^{\mathrm{TR}}$, retention intersects the common support with branch $b$'s interval $I_b(\xi^{\mathrm{TR}})$. If the public mean jump after that retention event is $\Delta m_b^{\mathrm{ret}}(\xi^{\mathrm{TR}})$, then
\[
  K^+=\bigl(K\cap I_b(\xi^{\mathrm{TR}})\bigr)-\Delta m_b^{\mathrm{ret}}(\xi^{\mathrm{TR}}),
  \qquad
  \bar g^+=\bar g-\Delta m_b^{\mathrm{ret}}(\xi^{\mathrm{TR}}),
  \qquad
  q^+=q.
\]
Terminal disclosure by either branch resolves the episode and exits the unresolved state space. Thus unresolved histories use only retention restrictions, and the two-route application is evaluated on an exact common gap Markov kernel.
\end{proposition}

\begin{proposition}[Certified-copy race verification]\label{prop:race_verification}
Suppose branch review clocks are public conditional on the announced race state, have no simultaneous jumps, and are independent of hidden content conditional on that state. At a review jump only the active branch chooses. Fix interval race policies and locally bounded continuation values $V^\ell$, $\ell\in\{0,i,j\}$, satisfying transversality, the branchwise race Bellman equations, and the one-step retention/terminal-disclosure inequalities at every branch review slice. Then the race profile is a pure MPBE on the certified-copy race state space, where $\mathcal M_b$ terminal disclosure by branch $b$ and $\mathcal K_b$ branch-$b$ retention with the common-gap censoring update.
\end{proposition}

\begin{proposition}[Cross branch censoring from an observed retention event]\label{prop:cross_branch_censoring}
Fix an unresolved public slice $\xi^{\mathrm{TR}}$ in the two-route race. Suppose branch $j$ retains at a public review date whenever $g\in I_j(\xi^{\mathrm{TR}})$ for some interval $I_j(\xi^{\mathrm{TR}})$. Then an observed retention on branch $j$ updates the common gap support to
\[
  K^{j,\mathrm{ret}}(\xi^{\mathrm{TR}})
  =
  \bigl(K\cap I_j(\xi^{\mathrm{TR}})\bigr)-\Delta m_j^{\mathrm{ret}}(\xi^{\mathrm{TR}}).
\]
Since branch $i$ carries the same certified record, branch $i$'s feasible post-event gap set is exactly $K^{j,\mathrm{ret}}(\xi^{\mathrm{TR}})$ on the updated unresolved slice. Hence, a public retention event on branch $j$ is immediately informative about branch $i$'s unresolved private state even before branch $i$ next moves. The rival branch's feasible set can tighten or truncate purely through exact common gap censoring.
\end{proposition}

\begin{corollary}[Waiting set shrinkage under cross-slice stability]\label{cor:cross_branch_shrinkage}
In the setting of Proposition~\ref{prop:cross_branch_censoring}, suppose now the retained common gap set removes an initial segment of branch $i$'s feasible states in branch $i$'s one-sided order. Let $\Delta_i^{-}(d,\xi^{\mathrm{TR}})$ denote branch $i$'s pre-event next review disclosure gain and $\Delta_i^{+}(d^+,\xi^{\mathrm{TR},+})$ its post-event gain on the updated slice. If the correspondence between pre and post-event branch $i$ gaps satisfies the cross-slice dominance condition
\[
  \Delta_i^{+}(d^+,\xi^{\mathrm{TR},+})
  \ge
  \Delta_i^{-}(d,\xi^{\mathrm{TR}})
\]
for every surviving state and both gains are order preserving in the same one-sided order on the relevant slices, then the post-event waiting region of branch $i$ is weakly smaller than the pre-event waiting region restricted to the surviving support.
\end{corollary}

A primitive sufficient condition for cross-slice dominance is increasing differences of branch $i$'s disclosure gain in the gap and public slice, together with a monotone-likelihood-ratio improvement of the post-retention law, provided the shape improvement uniformly dominates the common public-mean recentering of survivng gaps.

\subsection{Independent record counterfactual}\label{subsec:network_independent_counterfactual}

\begin{corollary}[Common gap censoring versus an independent gap counterfactual]\label{cor:no_common_gap_counterfactual}
Maintain all hypotheses of Corollary~\ref{cor:cross_branch_shrinkage}. Compare the exact common gap race with a counterfactual race in which branch $j$'s retention is observed but branch $i$'s unresolved private gap has an independent hidden component. Both experiments condition on the same public branch $j$ audit mark, reputation update, and route status. The comparison varies whether branch $i$'s support is also restricted by branch $j$'s retention. Now suppose the common public update from branch $j$'s retention leaves branch $i$'s independent gap waiting region weakly larger than branch $i$'s pre-event waiting region. Then branch $i$'s post-retention waiting region in the common gap race is weakly contained in its post-retention waiting region in the independent gap counterfactual. The containment is moreover strict on any positive probability slice on which branch $j$'s retention removes positive conditional mass from branch $i$'s one-sided lower tail, and the disclosure gain comparison is strict on the surviving boundary.

Equivalently, let $\xi^{\mathrm{com},+}_{i|j}$ be the common-gap post-retention state, in which branch $i$'s support is restricted by branch $j$'s retention, and let $\xi^{\mathrm{ind},+}_{i|j}$ be the independent-record post-state with the same public branch-$j$ audit mark, reputation update, route status, and public mean jump, but with branch $i$'s hidden support updated only by its own independent-record law. Define
\[
  \Theta_0^{\mathrm{post}}(\xi^{\mathrm{com},+}_{i|j},\xi^{\mathrm{ind},+}_{i|j})
  :=
  V_0^{\mathrm{com}}(\xi^{\mathrm{com},+}_{i|j})
  -
  V_0^{\mathrm{ind}}(\xi^{\mathrm{ind},+}_{i|j}).
\]
This value is nonnegative under the corollary's order and dominance hypotheses, and is positive under the strict positive mass condition. The corresponding pre-event value at slice $\xi^{\mathrm{TR}}$ is
\[
  \Theta_0^{\mathrm{pre}}(\xi^{\mathrm{TR}})
  :=
  p_j^{\mathrm{ret}}(\xi^{\mathrm{TR}})\,
  \Theta_0^{\mathrm{post}}(\xi^{\mathrm{com},+}_{i|j},\xi^{\mathrm{ind},+}_{i|j}),
\]
or the analogous integral when the retained slice is stochastic.
\end{corollary}

\begin{corollary}[Censoring-only network value]\label{cor:censoring_only_network_value}
In the two-route certified-record race, choose primitives with $0$ route-credit rents and normalize the independent-record direct-access, racing, assignment, and cost terms to $0$, or below any prescribed $\varepsilon>0$. If the hypotheses of Corollary~\ref{cor:no_common_gap_counterfactual} hold and $\E[\Theta_0^{\mathrm{pre}}]>0$, the common-record censoring component of planner value is strictly positive; when the residual terms are below that component, completing the parallel certified branch strictly raises $W^P$.
\end{corollary}

These race results are the direct analogues of the path recursion: both branches carry the same docketed record, so the common gap representation preserves the single-file logic while allowing routes to compete for first terminal disclosure. Corollary~\ref{cor:censoring_only_network_value} is actually the tether to network formation: removing the common hidden gap removes the cross-branch censoring value, leaving only direct access and preemption forces.

\subsection{Direct access, timing, and formation inputs}\label{subsec:network_optimal}

The network force isolated above is censoring of the same docketed record across certified routes. The formation theorem therefore evaluates a new certified branch by decomposing its planner value into the direct access value it would have if it carried an independent record, the change in future assignment value, and the additional value $\Delta\Theta_0^j$ created by the shared record. Access, preemption, and credit for being first enter as timing primitives, not as a separate formation force; a Poisson specification extension is collected in the supplementary Online Appendix OB.

\subsection{Planner and private formation values}\label{subsec:network_formation_values}

The formation benchmark distinguishes productive future assignment value $A_i(G)$ from private route credit rent $C_i(G)$. The first captures future use of public reputation in productive assignments; the second captures private visibility or business stealing rents from sitting on the route that resolves first. Our main welfare object is
\begin{equation}
  W^{P}(G)
  =
  V_0(G)
  +
  \sum_{i\in N\setminus\{0\}} A_i(G)
  -
  \sum_{(i,j)\in L} c_{ij}
  -
  \sum_{(j,0)\in L} r_{j0}^{\mathrm{rec}},
  \label{eq:network_social_value}
\end{equation}
where $c_{ij}\ge 0$ is the present value maintenance cost of link $(i,j)$ and $r_{j0}^{\mathrm{rec}}\ge 0$ is the decision maker's reception or processing cost of maintaining incoming link $(j,0)$. This benchmark counts decision-maker value, productive future assignment, and maintenance costs; route-credit rents enter private payoffs. Remark that when a new parallel branch $j$ is added beside an incumbent branch $i$, the planner's productive assignment term includes both the new branch's assignment gain $\Delta A_j$ and the incumbent branch's cannibalized assignment value $\Delta A_i<0$, because the incumbent is less likely to be first after the race is opened. Weighted criteria $W^\omega:=W^P+\omega\sum_{i\neq0}V_i^{\mathrm{mat}}(G)$ can be handled for any fixed $\omega$ whenever the corresponding strict cost interval is nonempty.

By contrast, a non-DM agent $i$ evaluates network formation through
\begin{equation}
  \tilde V_i(G)
  =
  V_i^{\mathrm{mat}}(G)
  +
  \phi_i A_i(G)
  +
  C_i(G)
  -
  \sum_{j:(i,j)\in L} c_{ij},
  \qquad \phi_i\in[0,1].
  \label{eq:private_formation_value}
\end{equation}
Links into the DM require separate consent by the DM. Let $r_{j0}^{\mathrm{rec}}\ge 0$ denote the DM's private reception or processing cost from maintaining link $(j,0)$. Her private link consent payoff is
\begin{equation}
  \tilde V_0(G)=V_0(G)-\sum_{(j,0)\in L}r_{j0}^{\mathrm{rec}}.
  \label{eq:dm_private_link_value}
\end{equation}
The wedge between planner value in \eqref{eq:network_social_value} and the private payoff profile in \eqref{eq:private_formation_value}--\eqref{eq:dm_private_link_value} is the source of Proposition~\ref{prop:network_inefficiency}. The same logic applies to any fixed weighed welfare criterion for which the strict cost interval in the formation proposition is nonempty.

\begin{remark}[Link consent and transfers]\label{rem:transfers_route_credit}
The benchmark allows no transfers: a link forms only with consent from both endpoints. Equivalently, allowing transfers tied to link formation or terminal disclosure leaves the benchmark unchanged as long as those transfers cannot price the continuation value created when retention on one certified branch changes beliefs on another branch carrying the same record. If agents could bargain multilaterally over that value, the failure to form the branch identified above would disappear. In the supplementary Online Appendix OB, I note the mirror case in which credit for being on the successful route is a private rent and may instead generate excess links.
\end{remark}

\begin{definition}[Disclosure-closed feasible set]\label{def:disclosure_closed_lmax}
A feasible link set $L^{\max}$ is disclosure-closed for the formation exercise if every disclosure game generated by a network $G\subseteq L^{\max}$, and by each addition or deletion of a single link used to test pairwise stability, remains in the class analyzed here. Specifically, each operative component has either no active route, a path carrying one operative record and satisfying the hypotheses for interval policies, or a race between two certified routes carrying the same docketed record and satisfying the censoring hypotheses used below. If no active route exists, the continuation value is that of the same public state with no pending record.
\end{definition}

In the four-node benchmark $L^{\max}=\{(e,i),(i,0),(e,j),(j,0)\}$, disclosure closure is immediate: one-link tests produce either the incumbent path, the completed $j$ path, the completed two-route race, or a partial branch with no operative path to the decision maker.

\subsection{Formation in the race benchmark}\label{subsec:network_formation}

I next study bilateral consent formation following \citet{JacksonWolinsky1996}. The incumbent direct route $e\to i\to0$ is present; the question is whether the complementary links $e\to j$ and $j\to0$ remain absent, though together they activate a socially valuable parallel certified branch. The main-text wedge is the common-record censoring value created by the completed branch.

Let $L^{\max}\subseteq N\times N$ denote the feasible directed links. Forming $(i,j)$ requires consent of both endpoints. The sender $i$ bears maintenance cost $c_{ij}\ge 0$, and if $j=0$ the decision maker also bears reception cost $r_{i0}^{\mathrm{rec}}$ from \eqref{eq:dm_private_link_value}. Once a network $G=(N,L)$ is realized, the single-file path or two-route disclosure game is played and agents receive the private values in \eqref{eq:private_formation_value} and \eqref{eq:dm_private_link_value}.

In what follows, I use two standard concepts.

\begin{definition}[Pairwise stability]\label{def:pairwise_stability}
A network $G$ is \emph{pairwise stable} if:
\begin{enumerate}[label=(\roman*),nosep]
  \item for every link $(i,j)\in L$, neither endpoint strictly gains by severing it:
  \[
    \tilde V_i(G) \ge \tilde V_i\bigl(G-(i,j)\bigr),
    \qquad
    \tilde V_j(G) \ge \tilde V_j\bigl(G-(i,j)\bigr);
  \]
  \item no nonlink $(i,j)\in L^{\max}\setminus L$ has both endpoints weakly gaining, and at least one strictly gaining, from adding $(i,j)$.
\end{enumerate}
Here $\tilde V_0$ is defined by \eqref{eq:dm_private_link_value} and $\tilde V_i$ for $i\neq 0$ is defined by \eqref{eq:private_formation_value}.
\end{definition}

\begin{definition}[Efficiency]\label{def:efficiency}
A network $G$ is \emph{efficient} if it maximizes the planner-centric value $W^P(G)$ in \eqref{eq:network_social_value} among all feasible networks.
\end{definition}

The central decomposition separates independent-record value from common-record censoring. In the two-branch calculations write
\[
  \Delta V_0^j=
  \underbrace{V_0^{\mathrm{ind}}(G^1)-V_0(G^0)}_{\Delta V_0^{j,\mathrm{ind}}}
  +
  \underbrace{\bigl[V_0^{\mathrm{com}}(G^1)-V_0^{\mathrm{ind}}(G^1)\bigr]}_{\Delta\Theta_0^j},
\]
where both terms are integrated over the same docket-date public slice distribution. Completing branch $j$ changes planner value by
\[
  \Delta W^P
  =
  \Delta V_0^{j,\mathrm{ind}}+\Delta\Theta_0^j+
  \Delta A_j+
  \Delta A_i
  -c_{ej}-c_{j0}-r_{j0}^{\mathrm{rec}},
  \qquad \Delta A_i<0.
\]

\begin{proposition}[Pairwise formation can miss common-record censoring value]\label{prop:network_inefficiency}
Suppose the feasible link set is disclosure-closed, Proposition~\ref{prop:path_ladder} and Theorem~\ref{thm:dynamic_unraveling} apply on every operative path, and the completed branch activates the certified-record two-route race of Corollary~\ref{cor:censoring_only_network_value}. Suppose also that branch $j$ is operative only when both links $(e,j)$ and $(j,0)$ are present, and that the strict common-record condition gives $\Delta\Theta_0^j>0$. There exist environments in which the incumbent path $G^0=\{(e,i),(i,0)\}$ is pairwise stable, yet the completed branch $G^1=G^0\cup\{(e,j),(j,0)\}$ strictly raises $W^P$.

In particular, if primitives and costs satisfy
\[
  \Delta V_0^{j,\mathrm{ind}}+\Delta A_j+\Delta A_i
  <
  c_{ej}+c_{j0}+r_{j0}^{\mathrm{rec}}
  <
  \Delta V_0^{j,\mathrm{ind}}+\Delta\Theta_0^j+
  \Delta A_j+\Delta A_i,
\]
then the branch is absent in the independent-record counterfactual and socially valuable in the common-record benchmark. Note that each missing link is privately useless alone: adding only $(e,j)$ gives branch $j$ no operative route to the decision maker, while adding only $(j,0)$ gives $j$ no record access and imposes the receiver cost on the decision maker. The planner gain appears only when the two links jointly activate the certified branch and its common-record censoring value. With independent-record, assignment, and route-credit terms normalized to $0$, the interval reduces to $0<c_{ej}+c_{j0}+r_{j0}^{\mathrm{rec}}<\Delta\Theta_0^j$, so the formation failure is exactly that common-record censoring wedge. The same strict underconnection conclusion holds for any fixed weighted benchmark $W^\omega$ whose corresponding strict interval is nonempty.
\end{proposition}

\begin{corollary}[Censoring driven planner wedge]\label{cor:censoring_only_underconnection}
In the formation benchmark, normalize $\Delta V_0^{j,\mathrm{ind}}$ to $0$, set route-credit rents to $0$, and choose costs inside the strict interval in Proposition~\ref{prop:network_inefficiency}. Then the planner's strict gain from completing branch $j$ is generated entirely by $\Delta\Theta_0^j$.
\end{corollary}

Note that the mechanism is not ordinary link complementarity. For example, let branch $j$'s retention remove a lower tail of the common gap and recenter the public mean upward. If branch $i$'s next-review waiting set is $(-\infty,\zeta_i)$ on the surviving one-sided slice, the retention event shifts positive mass of branch-$i$ states out of waiting and into forward movement, raising $V_0$ by a strictly positive common-record increment. In the independent-record counterfactual the same branch-$j$ event leaves branch $i$'s support and cutoff unchanged, so this increment vanishes.

The same informational term appears when the two links are formed as a package. Normalize away any private rent from credit for the successful route, and let
\[
  S^{\mathrm{pkg}}:=\Delta V_0^{j,\mathrm{ind}}+\Delta A_j+\Delta A_i-c_{ej}-c_{j0}-r_{j0}^{\mathrm{rec}}
\]
be the planner surplus from the package before adding the continuation value created by the shared record. The sign of \(S^{\mathrm{pkg}}\) is not a consent test, since it includes the incumbent's loss \(\Delta A_i<0\). Let \(P_{e,j,0}^{\mathrm{pkg}}\) denote the aggregate private surplus of the endpoints, counting \(\Delta\Theta_0^j\) only when that value accrues to them or can be contracted over by them. If \(P_{e,j,0}^{\mathrm{pkg}}<0\) while \(S^{\mathrm{pkg}}+\Delta\Theta_0^j>0\), or if some endpoint's own consent constraint fails, the package is rejected even though the planner would add the branch. I collect the corresponding accounting variants, for completeness, in supplementary Online Appendix OB.

\section{Conclusion}\label{sec:conclusion}

Public delay is not neutral when hard evidence has a public life before disclosure. If the existence of a sealed record, its custodian, and its review date are observed, a decision not to advance the record rules out the states in which advancement would have occurred. The paper formalizes this idea as a censoring problem. The public signal gives a Gaussian reference likelihood; the public review history restricts its support; and, under interval strategies, the resulting posterior is summarized by finitely many endpoints. On the invariant benchmark domain, the same recursion verifies pure Markov equilibria with interval retention and implies resolution of old records in finite time under reputational discipline, on compact unresolved slices, with expectation and tail bounds beyond compact support.

The network implication follows from the same logic. Certified copies do not merely add speed or direct access. When two routes carry the same docketed record, retention on one route restricts the states still possible on the other before that route acts. Common record censoring then creates planner value that private link formation need not price. Disclosure policy and network design are therefore linked by the public architecture of the record.

\appendix

\section{Proofs of Main Results}\label{app:proofs}

%

These manuscript appendices contain the material needed to reproduce the main arguments. Appendix~\ref{app:proofs} proves the stated results. Appendix~\ref{app:diffusion_realoptions} develops the filtering, review-censoring, and state-recursion machinery behind the exact public posterior. Appendix~\ref{app:extensions} specifies the reputational and formation primitives used in the equilibrium and network results. Additional benchmarks and extensions are collected in supplementary Online Appendices OA-OC.

\subsection{Proofs for Section~\ref{sec:model}}\label{app:proofs_model}

\begin{proof}[Proof of Proposition~\ref{prop:audit_separation}]
Conditional on the public slice $\xi$, the prior joint density is proportional to
\[
  \phi(d;\bar d,q)\mathbf 1\{d\in K\}\,p(\theta\mid\xi).
\]
Retention is the event $d\in B_i(\xi)$, and the audit mark has likelihood $\ell_\theta(s\mid \mathrm{ret},\xi)$ independent of $d$ conditional on $(\mathrm{ret},\xi,\theta)$. Hence
\[
  p(d,\theta\mid \mathrm{ret},s,\xi)
  \propto
  \phi(d;\bar d,q)\mathbf 1\{d\in K\cap B_i(\xi)\}
  p(\theta\mid\xi)\ell_\theta(s\mid \mathrm{ret},\xi),
\]
which factorizes after normalization. The relay and terminal cases replace the retention event by the corresponding custody event.
\end{proof}

\begin{proof}[Proof of Proposition~\ref{prop:interval_policy_mpbe_verification}]
The exact state $Z=(S,d)$ is Markov by the state recursion in Section~\ref{sec:dynamic}. On path, public histories induce the censored Gaussian posterior by Bayes' rule. Off-path, Definition~\ref{def:interval_policy_assessment} assigns the ordered-cut weak limit of the same Bayesian update, so beliefs are specified after every public review history.

The Bellman equation and transversality imply optimality between review dates. At a review mark, the one-step inequalities make the prescribed action attain the maximum between $\mathcal M^iV^i$ and $\mathcal K^iV^i$ at every feasible realized gap; value matching handles finite frontiers and the tie-breaking convention. The decision maker chooses the posterior mean under quadratic loss. Thus, strategies are sequentially optimal and beliefs are consistent after every public history, giving a pure MPBE.
\end{proof}

\subsection{Proofs for Section~\ref{sec:dynamic}}\label{app:proofs_dynamic}
\begin{proof}[Proof of Theorem~\ref{thm:primitive_ou_quadratic_audit}]
Proposition~\ref{prop:state_compression} gives the exact Gaussian likelihood for the holder's gap after every public history, and Corollary~\ref{cor:endpoint_union_closure} gives finite interval-union support under interval policies. Assumption~\ref{assumption:compact_interval_benchmark} keeps the feasible slices inside the compact invariant domain and makes every review date strict waiting set an interval inside the current support. The ordered-cut completion in Definition~\ref{def:interval_policy_assessment} supplies beliefs at zero-probability review actions. The compact benchmark therefore constructs slice-specific intervals $I_i^\ast(\xi)$ and continuation values satisfying the Bellman equation and review date inequalities in Proposition~\ref{prop:interval_policy_mpbe_verification}. That proposition verifies the constructed assessment as a pure interval-policy MPBE.
\end{proof}

\begin{proof}[Proof of Proposition~\ref{prop:path_ladder}]
At each feasible review slice, Theorem~\ref{thm:primitive_ou_quadratic_audit} gives a retention set $K(\xi)\cap I_i^\ast(\xi)$ and forward movement on its complement. Proposition~\ref{prop:state_compression} and Proposition~\ref{prop:full_state_recursion} gives the exact endpoint update after each retention, relay, or terminal disclosure. Iterating over the finite path yields the stated ladder.
\end{proof}

\begin{proof}[Proof of Corollary~\ref{cor:audit_refresh_discipline}]
With $M_i(\pi)=(1-\chi_i)\pi+\chi_i\pi_i^0$, $\chi_i\in(0,1)$ and $\pi_i^0\in(0,1)$,
\[
  M_i([0,1])=[\chi_i\pi_i^0,1-\chi_i(1-\pi_i^0)]\subset(0,1).
\]
The support condition keeps all realized prompt and skipped-review posteriors in this compact interior interval. Bayes plausibility gives a common posterior mean, and the prompt experiment has variance at least $\eta_i$ above the skipped-review experiment. Since $G_i$ is the unweighted continuation contribution of reputation, its curvature already includes the relevant persistence and discounting. The second-order integral form of convexity gives an expected continuation-value gain at least $\gamma_i\eta_i/2$. Thus Assumption~\ref{assumption:uniform_discipline} holds with $\underline\delta_i=\gamma_i\eta_i/2$; the payoff-weighted loss used in Theorem~\ref{thm:dynamic_unraveling} is $\beta_i\underline\delta_i$.
\end{proof}

\begin{proof}[Proof of Theorem~\ref{thm:dynamic_unraveling}]
Fix an unresolved public history at which sender $i$ holds the operative file. On the maintained compact region, retaining at an active review has material gain at most $\overline G_i(a)$ by Assumption~\ref{assumption:material_age_envelope} and loses at least $\beta_i\underline\delta_i$ in reputational continuation value by Assumption~\ref{assumption:uniform_discipline}. For $a\ge a_i^\dagger$, the definition of $a_i^\dagger$ gives $\overline G_i(a)\le\beta_i\underline\delta_i$, so retention cannot strictly improve the continuation value. The tie convention selects relay or terminal disclosure.

If the file remains outstanding at $i$ until calendar date $\tau+a_i^\dagger$, review recurrence gives
\[
  \int_{\tau+a_i^\dagger}^{\infty}\widetilde\lambda_u^i\,du=\infty
  \qquad \text{a.s.}
\]
The conditional survival probability of no further active review after that date is the expectation of the exponential of the negative integrated intensity and therefore converges to $0$. Thus the first active review after the threshold occurs in finite time almost surely and moves the file forward. On the maintained compact-slice event, finite downstream custody gives terminal disclosure in finite calendar time almost surely.
\end{proof}

\begin{proof}[Proof of Corollary~\ref{cor:gaussian_tail_old_record}]
At an active review, the incremental value of retention is bounded above by material gain minus $\beta_i\underline\delta_i$. Taking conditional expectations gives
\[
  \E[\text{retention gain}\mid \xi,a]
  \le \overline G_i^E(a)-\beta_i\underline\delta_i,
\]
which is nonpositive for $a\ge a_i^E$. Under the stated tail bound, pointwise profitability can occur only on states where $G_i^{\mathrm{mat}}(d,\xi;\varrho)>\beta_i\underline\delta_i$, an event of probability at most $\varepsilon_i(a)$.
\end{proof}

\subsection{Proofs for Section~\ref{sec:network}}\label{app:proofs_network}

\begin{proof}[Proof of Proposition~\ref{prop:race_verification}]
By Proposition~\ref{prop:race_full_state_recursion}, the race admits a Markov state. Since review jumps are never simultaneous, each review assigns the move to a single active branch. Between reviews, transversality and the Bellman equations for the relevant continuation values imply optimality. When branch \(b\) is reviewed, the review inequalities make the prescribed action attain the maximum among the admissible actions at that state. Beliefs follow Bayes' rule for the common gap along the equilibrium path; after histories assigned probability zero, they are completed by the rule based on the ordered cut. The assessment is therefore a pure Markov perfect Bayesian equilibrium. No preemption can arise between reviews, since no action is admissible before the next review jump.
\end{proof}

\begin{proof}[Proof of Proposition~\ref{prop:cross_branch_censoring}]
Fix an unresolved public slice $\xi^{\mathrm{TR}}$.
By Lemma~\ref{lem:race_state}, both branches carry certified copies of the same docketed record and their branch gaps equal the common hidden gap, $d_i=d_j=g$.
An observed retention on branch $j$ is therefore exactly the public event
\[
  g\in I_j(\xi^{\mathrm{TR}}).
\]
Bayes' rule intersects the current common support $K$ with this interval.
The public mean jump after retention translates the common gap by $-\Delta m_j^{\mathrm{ret}}(\xi^{\mathrm{TR}})$, so Proposition~\ref{prop:race_full_state_recursion} gives
\[
  K^{j,\mathrm{ret}}(\xi^{\mathrm{TR}})
  =
  \bigl(K\cap I_j(\xi^{\mathrm{TR}})\bigr)-\Delta m_j^{\mathrm{ret}}(\xi^{\mathrm{TR}}).
\]
Since branch $i$'s unresolved gap is the same common gap on the updated slice, branch $i$'s feasible post-event set is exactly this support.
Thus public retention on branch $j$ is immediately informative for branch $i$ because both branches are written on the same hidden scalar. The rival feasible set can tighten or truncate through exact common gap censoring.
\end{proof}

\begin{proof}[Proof of Corollary~\ref{cor:cross_branch_shrinkage}]
Let $W_i^{-}$ be the pre-event waiting region restricted to the surviving support and let $W_i^{+}$ be the post-event waiting region. Take any post-event feasible gap $d^+\in W_i^{+}$ and let $d$ be its corresponding pre-event branch-$i$ gap under the common public mean translation. Since $d^+$ is in the post-event waiting region, $\Delta_i^{+}(d^+,\xi^{\mathrm{TR},+})<0$. By cross-slice dominance,
\[
  \Delta_i^{-}(d,\xi^{\mathrm{TR}})
  \le
  \Delta_i^{+}(d^+,\xi^{\mathrm{TR},+})
  <0.
\]
Thus $d$ belongs to the pre-event waiting region on the surviving support. The preceding display gives a pointwise inclusion: every post-event waiting state has a pre-event antecedent that is also a waiting state. Applying this pre and post-event correspondence to the pointwise inclusion yields the claimed weak set inclusion.
\end{proof}

\begin{proof}[Proof of Corollary~\ref{cor:no_common_gap_counterfactual}]
In both experiments the public observes branch $j$'s retention, the same audit mark, the same reputation update, the same public mean jump, as well as the same route status update. These public coordinates are held fixed in the comparison. The common post-state $\xi^{\mathrm{com},+}_{i|j}$ restricts branch $i$'s support by branch $j$'s retention; the independent post-state $\xi^{\mathrm{ind},+}_{i|j}$ keeps branch $i$'s hidden component at its independent-record support after the same public coordinates have been applied.

In the common gap race, Proposition~\ref{prop:cross_branch_censoring} intersects the common support with branch $j$'s retention interval and applies the public mean translation. Under the lower tail removal and cross-slice dominance hypotheses of Corollary~\ref{cor:cross_branch_shrinkage}, this operation weakly shrinks branch $i$'s waiting region relative to the surviving pre-event support. In the independent gap counterfactual, branch $i$'s support is unchanged by branch $j$'s hidden file event. The maintained competition condition makes the independent gap waiting region weakly larger than the pre-event waiting region after the common public update from branch $j$'s retention. This means that the common gap post-retention waiting region is weakly contained in the independent gap post-retention waiting region.

If the retention event removes positive conditional mass from branch $i$'s one-sided lower tail and the disclosure gain comparison is strict at the surviving boundary, the monotone waiting set comparison removes a positive mass subset of states that would still be waiting states in the independent gap benchmark. The containment is then strict.

The continuation value difference defining $\Theta_0^{\mathrm{post}}(\xi^{\mathrm{com},+}_{i|j},\xi^{\mathrm{ind},+}_{i|j})$ is nonnegative because the common gap transition either leaves the DM's review value unchanged or moves states from waiting to disclosure while preserving the same subsequent public kernel on the surviving slice. Strict waiting set shrinkage on a positive probability slice yields a strictly positive contribution to the DM's continuation value. Multiplying by the pre-event probability of branch $j$'s retention, or integrating over retained slices, gives $\Theta_0^{\mathrm{pre}}$.
\end{proof}

\begin{proof}[Proof of Corollary~\ref{cor:censoring_only_network_value}]
By Corollary~\ref{cor:no_common_gap_counterfactual}, $\Theta_0^{\mathrm{pre}}$ is the expected present value of the extra continuation value for the decision maker, at the relevant slice before the event, generated when retention on one certified route restricts feasible states on the other. With rents from route credit set to $0$, and with the direct access, racing, assignment, and cost terms for separate records normalized away, $\E[\Theta_0^{\mathrm{pre}}]$ is the planner contribution of the shared record. If the remaining value with separate records is bounded by $\varepsilon$, the strict conclusion follows whenever $\E[\Theta_0^{\mathrm{pre}}]>\varepsilon$.
\end{proof}

\begin{proof}[Proof of Proposition~\ref{prop:network_inefficiency}]
Use the structural Poisson/two-route subclass of Appendix~\ref{app:extensions} and
\[
  L^{\max}=\{(e,i),(i,0),(e,j),(j,0)\}.
\]
The incumbent route $e\to i\to0$ is present and its two link costs are chosen small enough that incumbent severance is never profitable. In $G^0=\{(e,i),(i,0)\}$, adding only $(e,j)$ gives $j$ no operative path to the decision maker and costs $e$; adding only $(j,0)$ gives $j$ no record access and imposes the receiver cost on the decision maker. Thus $G^0$ is pairwise stable.

Completing branch $j$ creates decision-maker gain
\[
  \Delta V_0^j=\Delta V_0^{j,\mathrm{ind}}+\Delta\Theta_0^j,
\]
where the common-record component $\Delta\Theta_0^j$ is positive by Corollary~\ref{cor:censoring_only_network_value} on the stated primitive class. Productive assignment changes by $\Delta A_j+\Delta A_i$, with $\Delta A_i<0$. Choose primitives so that the total planner gain net of link and reception costs is positive, while each individual missing link remains privately unprofitable when added alone. Note this is possible on an open set because the individual one-link deviations activate no operative branch, whereas the two-link addition activates the certified race and its common-record censoring value. Hence $G^0$ is pairwise stable and $W^P(G^1)>W^P(G^0)$.

The displayed cost inequalities in the proposition make the censoring-only comparison explicit. They make the independent-record planner gain negative and the common-record planner gain positive after adding $\Delta\Theta_0^j$. For any fixed weighted welfare criterion $W^\omega$, the same construction applies whenever the analogous strict interval remains nonempty.
\end{proof}

\begin{proof}[Proof of Corollary~\ref{cor:censoring_only_underconnection}]
With $\Delta V_0^{j,\mathrm{ind}}=0$ and route-credit rents set to $0$, the strict interval in Proposition~\ref{prop:network_inefficiency} reduces the planner gain from completing the branch to the positive common-record term $\Delta\Theta_0^j$ net of costs and assignment effects. Pairwise stability of the incomplete network is the same one-link argument used in the proposition.
\end{proof}


\section{Filtering, review censoring, and exact state recursions}\label{app:diffusion_realoptions}

This appendix records the scalar latent-gap filter, finite endpoint closure under interval policies, affine support transport between reviews, and the review date verification lemma. Public review dates turn retention into censoring: the posterior is a Gaussian likelihood kernel conditioned on feasible support, and interval policies make that support a finite interval union.

\subsection{Public-signal benchmark and censored Gaussian recursion}\label{app:diffusion_prelim}

In the benchmark of Assumption~\ref{assumption:markov_rep}, the fundamental follows the Ornstein-Uhlenbeck diffusion
\begin{equation}
  dX_t = \kappa(\bar x-X_t)dt + \sigma_X dB_t,
  \label{eq:app_diffusion}
\end{equation}
and the public signal is
\begin{equation}
  dY_t = hX_tdt + \sigma_Y dW_t^Y,
  \label{eq:app_public_signal}
\end{equation}
with $B$ and $W^Y$ independent.
The uncensored signal likelihood is the Kalman-Bucy reference filter:
\begin{align}
  d\tilde m_t &= \kappa(\bar x-\tilde m_t)dt + \tilde K_t\bigl(dY_t-h\tilde m_tdt\bigr), \label{eq:kalman_mean}\\
  \dot{\tilde v}_t &= \sigma_X^2 - 2\kappa \tilde v_t - h^2\sigma_Y^{-2} \tilde v_t^2,
  \label{eq:kalman_var}
\end{align}
where $\tilde K_t=h\sigma_Y^{-2}\tilde v_t$ is the nominal Kalman gain.
These equations give the uncensored signal likelihood coordinate. For a docketed file born at $\tau$, the gap likelihood parameters $(\bar d_t,q_t)$ come from the Gaussian smoother for $X_\tau$ conditional on the realized public signal path, mapped forward to the holder's file-based posterior mean and differenced from the public mean. Review histories condition this one-dimensional likelihood on the feasible support $K_t$. Thus exact posterior moments are recovered from $(\bar d_t,q_t,K_t)$: public retention truncates support and public relay can create holes, while $\tilde v_t$ remains the uncensored likelihood-coordinate variance.

\begin{lemma}[Closed smoother law for the gap reference]\label{lem:closed_smoother_law}
Let
\[
  \widehat x_{\tau\mid t}=\E[X_\tau\mid Y_{[0,t]}],\qquad
  P_{\tau\mid t}=\Var(X_\tau\mid Y_{[0,t]}),\qquad
  C_{\tau t}=\Cov(X_\tau,X_t\mid Y_{[0,t]}).
\]
In the uncensored Ornstein-Uhlenbeck public signal reference system,
\[
  d\widehat x_{\tau\mid t}
  =h\sigma_Y^{-2}C_{\tau t}\bigl(dY_t-h\tilde m_tdt\bigr),
\]
\[
  \dot P_{\tau\mid t}=-h^2\sigma_Y^{-2}C_{\tau t}^2,
  \qquad
  \dot C_{\tau t}=-\kappa C_{\tau t}-h^2\sigma_Y^{-2}C_{\tau t}\tilde v_t.
\]
Moreover,
\[
  \E[X_t\mid X_\tau,Y_{[0,t]}]
  =\tilde m_t+\frac{C_{\tau t}}{P_{\tau\mid t}}
    \bigl(X_\tau-\widehat x_{\tau\mid t}\bigr).
\]
Thus the Gaussian reference law for the current holder's gap is obtained from the closed smoother block $\mathcal H_t^\tau$, while review histories enter by conditioning that law on the support $K_t$ and by the public mean translations in Proposition~\ref{prop:state_compression}.
\end{lemma}

\begin{proof}[Proof of Lemma~\ref{lem:closed_smoother_law}]
Apply the Kalman-Bucy equations to the augmented linear state $(X_\tau,X_t)$ with observation row $(0,h)$. They give the displayed dynamics for $(\widehat x_{\tau\mid t},P_{\tau\mid t},C_{\tau t})$; the bivariate Gaussian regression formula gives the last display. Review histories then condition this uncensored likelihood block through the induced support set. Review-date mean jumps are censored moments of the pre-jump uncentered gap law, after which the announced gap is recentered.
\end{proof}

\begin{lemma}[Ornstein-Uhlenbeck decay after a one-time revelation]\label{lem:OU_decay}
Assume~\eqref{eq:app_diffusion}.
Fix $s<t$ and a $\sigma$-field $\mathcal G\subseteq\mathcal F_s$.
Then
\[
  \E[X_t\mid \mathcal G\vee\sigma(X_s)]-\E[X_t\mid \mathcal G]
  =
  e^{-\kappa(t-s)}\Bigl(X_s-\E[X_s\mid \mathcal G]\Bigr).
\]
In particular, revealing $X_s$ changes the conditional mean of $X_t$ by an amount that decays at rate $\kappa$.
Moreover,
\[
  \E\!\left[\bigl(\E[X_t\mid \mathcal G\vee\sigma(X_s)]-\E[X_t\mid \mathcal G]\bigr)^2 \middle| \mathcal G\right]
  = e^{-2\kappa(t-s)}\Var(X_s\mid\mathcal G).
\]
\end{lemma}

\begin{proof}
The Ornstein-Uhlenbeck Markov property gives
\[
  \E[X_t\mid \mathcal F_s]=\bar x + e^{-\kappa(t-s)}(X_s-\bar x),
\]
or, equivalently, the innovation after $s$ has conditional mean $0$ given $\mathcal F_s$. Taking conditional expectations of this identity w.r.t. $\mathcal G\vee\sigma(X_s)$ and $\mathcal G$ yields the first display. Squaring that display and conditioning on $\mathcal G$ gives the variance identity.
\end{proof}

\begin{lemma}[Affine transport of a censored Gaussian]\label{lem:tn_affine}
Let $Z$ be Gaussian with mean $\mu$ and variance $q>0$, conditioned on $Z\in K$, where $K\subseteq\R$ is a measurable set.
If $X=a+bZ$ for constants $a\in\R$ and $b\neq 0$, then $X$ is Gaussian with mean $a+b\mu$ and variance $b^2q$, conditioned on $X\in a+bK$.
If $K$ is a finite union of intervals, this specializes to the usual affine transport of a truncated or interval union censored normal.
\end{lemma}

\begin{proof}
A censored Gaussian is a Gaussian law conditioned on a measurable set $K$.
An affine transformation of the latent Gaussian variable therefore remains Gaussian, and the conditioning set is transformed affinely as well.
\end{proof}

\begin{lemma}[Support update under public relay]\label{lem:relay_preservation}
Fix a public relay from intermediary $i$ to downstream intermediary $j$ at time $t$ in the single-file review benchmark.
Suppose that immediately before the relay,
\[
  d_{t-}^i\mid \mathcal F_{t-}^p
  \sim
  \mathcal N(\bar d_{t-},q_{t-})\ \text{conditioned on}\ d_{t-}^i\in K_{t-},
\]
where $K_{t-}$ is a finite union of intervals.
Let $B_i(\xi)\subseteq K_{t-}$ be denote the measurable retention set used by the current pure policy on the current public slice, and let $\Delta m_t^{\mathrm{relay}}:=m_t-m_{t-}$ be the public mean jump induced by observing the relay.
Then $\mu_t^j=\mu_{t-}^i$ and
\[
  d_t^j = d_{t-}^i-\Delta m_t^{\mathrm{relay}}.
\]
Conditional on the public relay event,
\[
  d_t^j\mid \mathcal F_t^p
  \sim
  \mathcal N(\bar d_{t-}-\Delta m_t^{\mathrm{relay}},q_{t-})\ \text{conditioned on}\ d_t^j\in K_t^{\mathrm{relay}},
\]
where
\[
  K_t^{\mathrm{relay}} = \bigl(K_{t-}\cap B_i(\xi)^c\bigr)-\Delta m_t^{\mathrm{relay}}.
\]
In particular, if $K_{t-}$ is a single interval and $B_i(\xi)$ is a bounded interval, then $K_t^{\mathrm{relay}}$ is generically a two-component interval union.
\end{lemma}

\begin{proof}
The custody transfer is instantaneous and both intermediaries observe the same sealed file $X_\tau$.
Hence, their private posterior means coincide at the relay time, so the private-minus-public gap changes through the public mean jump triggered by observing the relay.
Because the relay is chosen exactly when the realized gap lies outside the retention set, the public conditioning event is $d_{t-}^i\in K_{t-}\cap B_i(\xi)^c$.
Translating by $-\Delta m_t^{\mathrm{relay}}$ yields the stated support update for the downstream gap.

\end{proof}

\begin{lemma}[Moment operators on interval unions]\label{lem:interval_union_moments}
Fix $q>0$, $\bar d\in\R$, and an ordered finite interval union
\[
  B=\bigcup_{r=1}^R [\ell_r,u_r],
  \qquad
  -\infty\le \ell_1\le u_1<\ell_2\le u_2<\cdots<\ell_R\le u_R\le \infty,
\]
where open, closed, or half-open endpoint conventions give the same formulas because the Gaussian reference law has no atoms.
Let $Z\sim \mathcal N(\bar d,q)$ under the reference kernel and write $\phi$ and $\Phi$ for the standard normal density and distribution function, with the conventions $\phi(\pm\infty)=0$, $\Phi(-\infty)=0$, $\Phi(\infty)=1$, and $\lim_{x\to\pm\infty}x\phi(x)=0$ (equivalently, $\pm\infty\cdot\phi(\pm\infty)=0$).
Define now standardized endpoints
\[
  \alpha_r:=\frac{\ell_r-\bar d}{\sqrt q},
  \qquad
  \beta_r:=\frac{u_r-\bar d}{\sqrt q},
\]
and interval union moment sums
\[
  M_0(B):=\sum_{r=1}^R \bigl[\Phi(\beta_r)-\Phi(\alpha_r)\bigr],
\]
\[
  M_1(B):=\sum_{r=1}^R \bigl[\phi(\alpha_r)-\phi(\beta_r)\bigr],
\]
\[
  M_2(B):=\sum_{r=1}^R \bigl[\alpha_r\phi(\alpha_r)-\beta_r\phi(\beta_r)+\Phi(\beta_r)-\Phi(\alpha_r)\bigr].
\]
If $M_0(B)>0$, then
\[
  \E[Z\mid Z\in B]=\bar d+\sqrt q\,\frac{M_1(B)}{M_0(B)},
\]
and
\[
  \Var(Z\mid Z\in B)=q\left(\frac{M_2(B)}{M_0(B)}-\left(\frac{M_1(B)}{M_0(B)}\right)^2\right).
\]
These moments are Borel measurable in $(\bar d,q,\ell_1,u_1,\dots,\ell_R,u_R)$ and are continuous on any admissible family for which $M_0(B)$ is bounded away from $0$.
In particular, for a review date event $B\subseteq K$, the static conditional mean operator is the measurable map
\[
  \mathfrak m(B;\bar d,q):=\E[Z\mid Z\in B],
\]
and the exact post-event conditional variance is a measurable function of $(\bar d,q,B)$. Central truncation can make it smaller than the reference kernel variance $q$; complement truncation can make it exceed $q$ when the event deletes enough central mass, as after relay conditioning on the complement of a central interval.
\end{lemma}

\begin{proof}
Integrating the Gaussian density over each component gives the standard truncated-normal mass and moment terms; summing components yields $M_0$, $M_1$, and $M_2$. The formulas are finite sums of continuous functions, with continuity whenever $M_0(B)$ is bounded away from $0$. The variance statement follows from the displayed expression.
\end{proof}

\begin{lemma}[Ordered-cut completion for off-path review beliefs]\label{lem:frontier_completion}
Fix a finite interval-union support $K$ and an atomless Gaussian reference kernel $\nu_{\bar d,q,K}$. An ordered action cut $\zeta=(e,s)$ selects an action side $C(\zeta)\subseteq K$. Any admissible sequence of interval event sets whose ordered frontiers converge to $\zeta$ and whose action side is $C(\zeta)$ has normalized Gaussian restrictions converging weakly to $\nu_{\bar d,q,K}(\cdot\mid C(\zeta))$ if this side has positive mass, and to the corresponding one-sided degenerate law $\delta_e$ if it collapses to a finite frontier point. Hence the endpoint and side pin down the off-path posterior.
\end{lemma}

\begin{proof}
For bounded continuous $f$, numerator and denominator of the normalized Gaussian integral converge by dominated convergence and atomlessness; if the side collapses, continuity gives convergence to $f(e)$.
\end{proof}

\begin{proposition}[Post-review kernels and interval union closure]\label{prop:kernel_markov_closure}
Encode every ordered interval union $K$ with at most $M$ conected components by its ordered endpoint vector.
On any admissible state space with at most $M$ components, affine translation of interval unions and the interval union moment formulas from Lemma~\ref{lem:interval_union_moments} imply that the review date maps
\[
  (\bar d,q,K)\mapsto (\bar d^{\mathrm{ret}},q^{\mathrm{ret}},K^{\mathrm{ret}})
  \qquad\text{and}\qquad
  (\bar d,q,K)\mapsto (\bar d^{\mathrm{relay}},q^{\mathrm{relay}},K^{\mathrm{relay}})
\]
are Borel measurable whenever the review date retention set $B_i(\xi)$ is itself a finite union of intervals.
Hence, once combined with the public state transition of $(m,\tilde v,\mathcal H,R,C,A)$, they induce Borel Markov kernels $Q_{\mathrm{ret}}^i$ and $Q_{\mathrm{relay}}^i$ on the announced path state.
Moreover, interval union support is preserved by both retention and relay.
\end{proposition}

\begin{proof}
Measurability of the update maps follows from three ingredients: measurable intersection and complement operations on finite unions of intervals at review dates, translation of ordered endpoints by the public mean jump, and measurable public mean jumps from Lemma~\ref{lem:interval_union_moments}.
A retention update intersects $K$ with another finite union of intervals and then translates the result.
A relay update intersects $K$ with the complement of a finite union of intervals and then translates the result.
Finite unions are preserved by both operations, so interval union support closure is exact in this class.
\end{proof}

\begin{corollary}[Linear component bound under interval policies]\label{cor:kernel_component_bound}
If, in addition, every review date policy interval $I_i(\xi)$ is a single interval, with actual retention event $K\cap I_i(\xi)$, then if $K$ has $N$ connected components immediately before a public review date, a retention update leaves at most $N$ components and a relay update leaves at most $N+1$ components.
Starting from $K_\tau=\R$, after $r$ public relays the support therefore has at most $1+r$ components.
Along a simple path $P=(e,i_1,\dots,i_k,0)$, one has $r\le k$, so the exact single-file recursion never requires more than $k+1$ connected components.
\end{corollary}

\begin{proof}
A retention update intersects $K$ with one interval and then translates the result, so it cannot create a new connected component.
A relay update intersects $K$ with the complement of one interval and then translates the result.
Because one connected interval can split at most one existing component of an ordered interval union, the relay update can increase the component count by at most one.
Inducting from the initial support $K_\tau=\R$ gives the bound $1+r$ after $r$ relays, and the path bound $k+1$ follows because a simple path has at most $k$ downstream relays before terminal disclosure.
\end{proof}

\begin{proof}[Proof of Proposition~\ref{prop:state_compression}]
At docketing the current holder's gap is Gaussian and $K_\tau=\R$. On an event-free custody segment $[s,t]$, write
\[
  \hat X_\tau(u)=\E[X_\tau\mid\mathcal F_u^p],
  \qquad
  \zeta_{\tau,u}=\frac{\Cov(X_u,X_\tau\mid\mathcal F_u^p)}{\Var(X_\tau\mid\mathcal F_u^p)}>0 .
\]
The linear-Gaussian projection gives
\[
  d_u^i=\zeta_{\tau,u}\bigl(X_\tau-\hat X_\tau(u)\bigr),
  \qquad
  d_t^i=\frac{\zeta_{\tau,t}}{\zeta_{\tau,s}}d_s^i
       +\zeta_{\tau,t}\bigl(\hat X_\tau(s)-\hat X_\tau(t)\bigr).
\]
Thus $d_t^i=\chi_{s,t}d_s^i+\psi_{s,t}$ with public $\chi_{s,t}>0$ and $\psi_{s,t}\in\R$, so Lemma~\ref{lem:tn_affine} yields $K_t=\chi_{s,t}K_s+\psi_{s,t}$. The support transports affinely, while $(\bar d_t,q_t)$ are recomputed from the current smoother likelihood.

At a public review date, retention is the event $d_{t-}^i\in K_{t-}\cap B_i(\xi)$ and relay is the complementary event inside $K_{t-}$. After retention,
\[
  K_t^{\mathrm{ret}}=(K_{t-}\cap B_i(\xi))-\Delta m_t^{\mathrm{ret}},
  \qquad
  \bar d_t^{\mathrm{ret}}=\bar d_{t-}-\Delta m_t^{\mathrm{ret}},
  \qquad q_t^{\mathrm{ret}}=q_{t-};
\]
relay is identical with $B_i(\xi)^c$ and $\Delta m_t^{\mathrm{relay}}$, as in Lemma~\ref{lem:relay_preservation}. Iterating event-free affine transport and review date conditioning gives a Gaussian likelihood kernel conditioned on the feasible support generated by public history. The payoff-relevant public history is therefore summarized by
\[
  S_t=(m_t,\tilde v_t,\mathcal H_t^\tau,R_t,C_t,A_t,\bar d_t,q_t,K_t)
\]
and the scalar hidden realization $d_t^{C_t}$.
\end{proof}

\begin{proof}[Proof of Corollary~\ref{cor:endpoint_union_closure}]
On the interval union equilibrium class, every review date retention set $B_i(\xi)$ is a finite union of intervals.
Between public events, affine transport preserves interval unions by Lemma~\ref{lem:tn_affine}.
At a retention review date, the updated feasible support is the intersection of the current interval union with another interval union, followed by translation, so it remains an interval union.
At a relay review date, the updated feasible support is the intersection of the current interval union with the complement of an interval union, again followed by translation, so it remains an interval union as well.
Encoding those components by ordered endpoints gives the finite endpoint representation stated in the corollary.
\end{proof}

\begin{proof}[Proof of Corollary~\ref{cor:endpoint_component_bound}]
By Corollary~\ref{cor:kernel_component_bound}, retention never increases the number of support components and each public relay increases that number by at most one when review date policy intervals are single intervals.
Starting from the single component $K_\tau=\R$, the exact path recursion therefore uses at most $1+r_t\le k+1$ components after $r_t$ relays on a path $P=(e,i_1,\ldots,i_k,0)$ with $k$ downstream custody transfers before the DM.
\end{proof}

\begin{proof}[Proof of Proposition~\ref{prop:full_state_recursion}]
Proposition~\ref{prop:state_compression} gives the exact scalar latent gap recursion, Proposition~\ref{prop:kernel_markov_closure} gives Borel post-review kernels for $(\bar d,q,K)$ on the interval union class, and Appendix~C supplies the public Markov law for reputations.
The reference coordinates $(\tilde v,\mathcal H)$ evolve by the Ornstein-Uhlenbeck public signal likelihood equations and the closed smoother law in Lemma~\ref{lem:closed_smoother_law}. The custody and age coordinates $(C,A)$ are public bookkeeping variables, and $R$ follows the public audit law. The public mean $m$ is tracked recursively by the smoother and by the review date jumps displayed in Proposition~\ref{prop:full_state_recursion}; these jumps are censored moments of the pre-jump uncentered gap law, after which the announced gap coordinate is recentered.
Between review dates these measurable updates combine into the transport functional $\Phi^i_{s,t}$ displayed in the main text, driven by the realized public signal segment on $[s,t]$ and with support component $K_t=\chi_{s,t}K_s+\psi_{s,t}$.
At a public review date, the support updates from Proposition~\ref{prop:state_compression} and the public updates of $(m,\tilde v,\mathcal H,R,C,A)$ combine into the measurable jump maps $T_i^{\mathrm{ret}}$, $T_{i\to j}^{\mathrm{relay}}$, and $T_i^{\mathrm{term}}(\widehat\xi,d)$ displayed in Proposition~\ref{prop:full_state_recursion}.
Taking the product of these between-review and review date transitions yields a Borel kernel for the full announced state
\[
  S_t=(m_t,\tilde v_t,\mathcal H_t^\tau,R_t,C_t,A_t,\bar d_t,q_t,K_t).
\]
Hence, $(S_t,d_t^{C_t})$ is Markov on the single-file benchmark, and the Bellman equation in the main text is written on an exact recursive law for the full announced state together with the current holder's scalar hidden gap.
\end{proof}

\begin{proof}[Proof of Lemma~\ref{lem:race_state}]
Certified copies condition on the same sealed record and public signal history, so the two branch gaps are the same scalar $g_t$. Between public events this scalar is transported by the single-file affine Gaussian recursion; branch retentions intersect its support with the active branch interval, branch-status variables enter $\Xi_t$, and terminal disclosure exits the unresolved state space. Thus the unresolved law is a Gaussian kernel for $g_t$ conditioned on a finite interval union $K_t$. Stacking branch gaps would only duplicate this scalar and produce a singular covariance matrix.
\end{proof}

\begin{proof}[Proof of Corollary~\ref{cor:race_union_closure}]
On each unresolved race slice, branch $b$'s waiting rule is an interval in the common hidden gap $g_t$. Hence, an unresolved retention event corresponds to intersecting the current common support $K_t$ with one interval. Terminal disclosure by either branch resolves the episode, so the unresolved support recursion ends at that event. Intersections with intervals preserve interval unions, and between public events the common support is transported affinely.
Docketing starts from the single interval $K_\tau=\R$. The expert's initial strategic relay to the two branches is itself a public policy event: when the expert relays only outside her retention interval, the common support is intersected with the complement of that interval and is generically a two-component interval union. Subsequent unresolved branch retentions intersect this finite union with branch policy intervals, and terminal disclosure exits the unresolved state space. Thus finite-union closure is part of the baseline two-route race.
\end{proof}

\begin{proof}[Proof of Proposition~\ref{prop:race_full_state_recursion}]
Lemma~\ref{lem:race_state} gives the exact common hidden state and Corollary~\ref{cor:race_union_closure} gives support closure. The remaining public coordinates use the same measurable path ingredients: reference likelihood equations, recursive public-mean tracking, censored review-date mean jumps, public audit law, elapsed age, and branch-status bookkeeping. Between reviews their product with affine support transport gives $\Phi^{\mathrm{TR}}_{s,t}$. At a branch-$b$ review, retention is $g\in I_b(\xi^{\mathrm{TR}})$, so $K$ becomes $(K\cap I_b(\xi^{\mathrm{TR}}))-\Delta m_b^{\mathrm{ret}}$, $\bar g$ is translated by the same jump, and $q$ is unchanged; terminal disclosure exits. All coordinates are Borel functions of the current unresolved slice, so $S_t^{\mathrm{TR}}$ is Markov.
\end{proof}

\subsection{Review verification}\label{app:review_control}

The dynamic problem in the main text is a public-review stopping problem.
The current custodian takes the review process as given and chooses whether to disclose or retain at each public review date.
In the next lemma I embed the review date analogue of the standard verification argument.

\begin{lemma}[Verification for public review control]\label{lem:verification_review}
Let $V:E\to\R$ be continuous and locally $C^2$ between review dates on an open state space $E$.
Suppose that for a current custodian $i$,
\[
  \rho V(z)
  =
  u(z)
  +
  \mathcal A V(z)
  +
  \lambda(S)\bigl(\max\{\mathcal M V(z),\mathcal K V(z)\}-V(z)\bigr)
\]
holds for Lebesgue-a.e. $z=(S,d)\in E$, where $\lambda(S)$ is public state measurable (and the usual transversality and integrability conditions hold).
Then $V$ dominates the payoff of every admissible review date policy.
If a policy chooses disclosure whenever $\mathcal M V\ge \mathcal K V$ and retention otherwise, that policy is optimal and $V$ is the value function.
\end{lemma}

\begin{proof}
Apply the generalized It\^o formula to $e^{-\rho t}V(Z_t)$ between review dates and add the review compensator. For any policy, its selected jump continuation $J_t^\alpha$ is bounded above by $\max\{\mathcal M V(Z_{t-}),\mathcal K V(Z_{t-})\}$, so the Bellman equation leaves a nonpositive residual drift after flow payoffs are accumulated. The discounted payoff process is therefore a supermartingale; transversality gives domination. If the policy attains the pointwise maximum at every review mark, the process is a martingale and the policy is optimal.
\end{proof}

\section{Reputational and formation primitives}\label{app:extensions}

The primitive objects used in the main text are specified here; ancillary derivations and illustrative special cases are collected, for completeness, in the supplementary Online Appendix OB.
The audit record is a public experiment separate from custody. If $\theta_i\in\{H,L\}$ is audit quality and $s$ is the audit mark after a review outcome, then $\pi_t^i=\pi_{t-}^i\ell_H(s)/[\pi_{t-}^i\ell_H(s)+(1-\pi_{t-}^i)\ell_L(s)]$ and $R_t^i=\rho_i(\pi_t^i)$. The custody outcome censors the sealed-file gap; the audit mark prices review quality. Public recertification $M_i(\pi)=(1-\chi_i)\pi+\chi_i\pi_i^0$, with $\chi_i,\pi_i^0\in(0,1)$, keeps audit beliefs interior. Corollary~\ref{cor:audit_refresh_discipline} uses the Bayes-plausible route on that interval: prompt resolution is uniformly more informative than skipped retention, and convex reputation value converts dispersion into a uniform audit-reputation advantage.

The formation section separates productive assignment value from route-credit rents. Productive assignment value is discounted surplus from future mandates awarded by public reputation; route credit is a private rent from occupying a salient position on the winning route. Planner welfare counts decision-maker value, productive assignment surplus, and link costs; private formation values add route-credit rents.

The dated-record Ornstein-Uhlenbeck primitive gives the material age envelope: on compact unresolved slices, revealing a record of public age $a$ changes future posterior means by an exponentially decaying bound, so quadratic material losses imply $G_i^{\mathrm{mat}}(d,\xi;\varrho)\le C_i e^{-\lambda_i^{\mathrm{age}}a}$. In the structural Poisson/two-route subclass used for the supplementary timing benchmark, downstream delay $T_j\sim\mathrm{Exp}(\bar\lambda_j)$, remaining disclosure value $\bar M e^{-\kappa a}$, and first-disclosure premium $\varpi_i$ yield $\Gamma_i(\xi)=\bar M e^{-\kappa a}(\rho+\kappa)/(\rho+\kappa+\bar\lambda_j)$, $\Pi_i(\xi)=\varpi_i$, and $\Lambda_i(\xi)=0$, the primitive lower bound for the direct-access comparison.



\singlespacing
\setlength{\bibsep}{0pt}
\bibliographystyle{abbrvnat}\bibliography{references}

\end{document}